\documentclass[journal]{IEEEtran}
\usepackage[ruled,algonl]{algorithm2e}
\usepackage{adjustbox}
\usepackage{multicol}
\usepackage{multirow}
\usepackage[table]{xcolor}
\usepackage{color}
{}
{}
{}
\usepackage{subcaption}
\usepackage{xcolor}

\graphicspath{{images/}}
\usepackage{float}
\usepackage{epstopdf} 
\usepackage{amsmath}
\usepackage{balance}
\usepackage{fontawesome}

\usepackage{booktabs,siunitx}
\usepackage{wrapfig}
\usepackage{amssymb}
\usepackage[T1]{fontenc}
\DeclareMathAlphabet{\mathpzc}{OT1}{pzc}{m}{it}
\usepackage{booktabs}
\usepackage{bbding}
\usepackage{tikz}
\def\checkmark{\tikz\fill[scale=0.4](0,.35) -- (.25,0) -- (1,.7) -- (.25,.15) -- cycle;} 

\renewcommand{\footnoterule}{%
\kern -3pt
\hrule width 0.49 \textwidth height 0.5pt
\kern 1pt
}

\begin{document}

\title{A Two-Stage Service Restoration Method for Electric Power Distribution Systems}
\author{Shiva Poudel,~\IEEEmembership{Student Member,~IEEE} and
        Anamika Dubey,~\IEEEmembership{Member,~IEEE,}
        \thanks{S. Poudel and A. Dubey are with the School of Electrical Engineering and Computer Science, Washington State University, Pullman, WA 99164, USA. E-mail: shiva.poudel@wsu.edu, anamika.dubey@wsu.edu}
    }
\maketitle

\begin{abstract}
   Improving the reliability of power distribution systems is critically important for both utilities and customers. This calls for an efficient service restoration module within a distribution management system to support the implementation of self-healing smart grid networks. The emerging smart grid technologies, including distributed generators (DGs) and remote-controlled switches, although enhance the self-healing capability and allow faster recovery, pose additional complexity to the service restoration problem, especially under cold load pickup (CLPU) conditions. In this paper, we propose a novel two-stage restoration framework to generate restoration solutions with a sequence of control actions. The first stage generates a restoration plan that supports both the traditional service restoration using feeder reconfiguration and the grid-forming DG-assisted intentional islanding methods. The second stage generates an optimal sequence of switching operations to bring the outaged system quickly to the final restored configuration. The problem is formulated as a mixed-integer linear program that incorporates system connectivity, operating constraints, and CLPU models. It is demonstrated using a multi-feeder test case that the proposed framework is effective in utilizing all available resources to quickly restore the service and generates an optimal sequence of switching actions to be used by the operator to reach the desired optimal configuration.
\end{abstract}
\maketitle


\section*{Nomenclature}
\addcontentsline{toc}{section}{Nomenclature}
\begin{IEEEdescription}[\IEEEusemathlabelsep\IEEEsetlabelwidth{$V_1,V_2,V_3$}]
    \item [{\textit{A. Sets}}] 
     \item [$\mathcal{E}_F$] Set of faulted or tripped switches  
     \item [$\mathcal{E}_S^t$] Set of normally-open tie switches 
     \item [$\mathcal{E}_S^s$] Set of normally-closed sectionalizing switches
     \item [$\Phi$ = $\{a, b, c\}$] Set of phases of a bus
    \item [$\mathcal{E}$] Set of physical lines
    \item [$\mathcal{V}$] Set of physical system buses
    \item [$\mathcal{V}_S$] Set of remotely switchable buses
    \item [$\mathcal{E}_S$] Set of switchable lines. i.e. $\mathcal{E}_S^r \cup \mathcal{E}_S^t$
    \item [$\mathcal{E}{_c}$] Set of switches in a cycle $c$ 
    \item [$\mathcal{E}_S^v$] Set of virtual edges for DGs connection  
    \item [$\mathcal{E}_{R}$] Set of voltage regulators 
    \vspace{0.5 cm}
 \item[{\textit{B. Variables}}]
    \item [$v_i$] Bus energization variable
    \item [$P_{e}^\phi+\text{j} Q_{e}^\phi$] Complex power flow from $i$ to $j$ 
    \item [$\delta_{e}$] Line or switch decision variable
    \item [$s_i$] Load pick-up variable
    \item [$\textbf{\textit{P}}_{e}+\text{j}\textbf{\textit{Q}}_{e}$] Three-phase complex power flow from $i$ to $j$
    \item [$\textbf{\textit{U}}_i$] Three-phase voltage magnitude square vector
    \item [$\textbf{\textit{V}}_i$] Three-phase voltage vector for bus $i$ 
    \vspace{0.5 cm}
\item [{\textit{C. Parameters}}] 
    \item [$\textbf{\textit{S}}_{e}^{rated}$] Apparent power flow limit for $e\in \mathpzc{E}$
    \item [$P_{Li}^\phi+\text{j} Q_{Li}^\phi$] Complex power demand at $i$ for phase $\phi \in \Phi$ 
    \item [$\textbf{\textit{S}}_{e}$] Polygon based linearized equivalent of $\textbf{\textit{S}}_{e}^{rated}$
    \item [$P_{G}^{max}$] Maximum active power capacity of a DG 
    \item [$Q_{G}^{max}$] Maximum reactive power capacity of a DG 
    \item [$|\mathcal{E}{_c}|$] Number of switches in a cycle $c$
    \item [$\textbf{r}_{e}$/$\textbf{x}_{e}$] Resistance/Reactance matrix of a line $e:(i,j) \in \mathpzc{E}$
    \item [$\textbf{\textit{P}}_{Li}+\text{j}\textbf{\textit{Q}}_{Li}$] Three-phase complex load demand at bus $i$
\end{IEEEdescription}

\section{Introduction}
The electric power grid is essential to modern society for economic prosperity, national security, and public health and safety. The electric power distribution systems are the ``last mile'' of the power supply that delivers electric power to the end-users. It is worth mentioning that 90\% of customer outage-minutes are due to the damages in the local distribution systems \cite{campbell2012weather}. Although several efforts have been made and to improve the distribution system reliability, the power outages and faults in the distribution systems are inevitable. Further, natural disasters can cause significant damages to the mid-voltage and low-voltage power distribution systems resulting in economic losses due to lost productivity and increased concerns to customer safety due to power disruptions for extended periods.  The increased dependence on electric power grid coupled with the increasing number of natural disasters motivates the need for improving the reliability and resilience of the electric power distribution systems \cite{report1, national2017enhancing}. 

In past few years, several efforts have been made towards developing self-healing power distribution systems that can automatically identify, isolate, and restore the outaged area to improve grid reliability, resilience, and security in response to a wide variety of emergencies such as natural calamities, intentional attack, or human error \cite{renz2010anticipates, amin2011smart}. The highest levels of service reliability are achieved when the distribution system is designed and operated to minimize the effects of any fault that may occur. Along these lines, the Fault Location Isolation and Service Restoration (FLISR) is one of the most critical applications currently being adopted by the majority of distribution utility companies and also made available by most advanced distribution management system (ADMS) vendors to autonomously manage the system outages\cite{adms2, dubey2020paving}. A distribution service restoration (DSR) algorithm is a part of the complete FLISR application that develops a service restoration plan in the aftermath of a disruption to restore the power supply to the healthy feeders using feeder reconfiguration and intentional islanding using distributed generation (DGs) resources. This paper is focused on developing computationally tractable algorithms for distribution service restoration (DSR) by leveraging all available resources including DGs while accounting for the critical connectivity and operating constraints.

\begin{figure*}[t]
    \centering
    \includegraphics[width=0.85\textwidth]{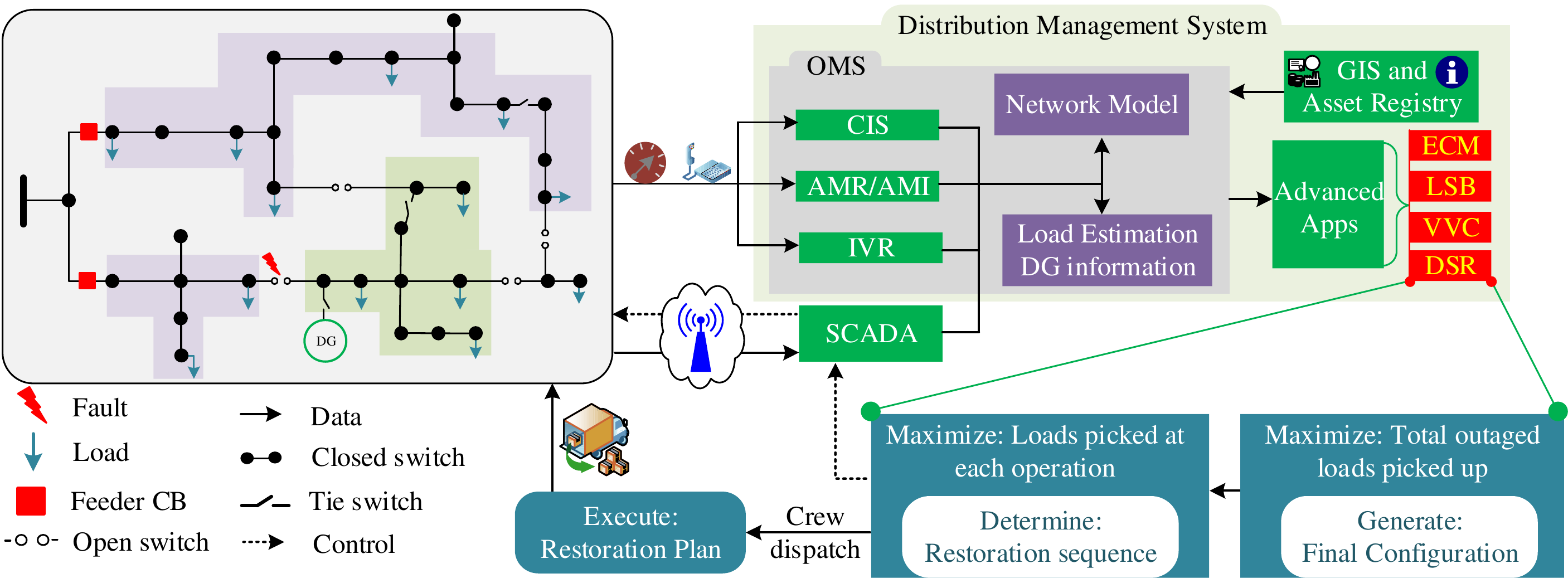}
    \caption{\small Proposed two-stage framework for distribution system restoration: work flow in a distribution management system (DMS)}
    \label{fig:1}
    \vspace{-0.4cm}
\end{figure*}

\vspace{-0.3 cm}

\subsection{Literature Review}
There is an extensive literature on solving the DSR problem for the power distribution systems. Earlier methods for DSR focused on designing expert systems, including heuristic search methods and use of soft-computing algorithms \cite{liu1988expert, miu2000multi, kumar2006service}. The use of feeder reconfiguration algorithms to restore the system for local outages has been well-studied in the literature, where DSR is normally modeled as a combinatorial optimization problem with operational and topological constraints \cite{zhou1997distribution, butler2004self, khushalani2007optimized}.  To address the growing concerns of natural disasters, recently, some emerging DSR methods have been proposed to restore distribution systems during natural disasters using intentional islanding supported by microgrids and DGs \cite{pham2009new,li2014distribution, wang2015self, chen2015resilient, sharma2015decentralized, chen2017multi, poudel2018critical}.

To reduce the outage duration, a DSR algorithm must (1) incorporate all available resources including DGs to restore a maximum number of customers and (2) generate an efficient restoration plan that allows the system operators to quickly restore the outaged system. Unfortunately, a majority of the existing DSR methods are limited to finding the final restored configuration for the given system. They formulate the DSR problem as a single-stage optimization problem. The single-step methods do not generate the sequence of intermediate feasible restored topologies that are required to reach the final restored configuration. Note that the system operators carry out a sequence of switching operations to restore the distribution system from a post-fault configuration to the final restored configuration. Thus, an efficient restoration plan should generate a sequence of switching actions required to bring the disrupted system to a restored network configuration that maximizes the total number of loads picked-up by using all available resources. Notice that the generation of switching sequence requires careful consideration of distribution system's operating constraints such that the intermediate topologies are power flow feasible and do not violate voltage and thermal limit constraints especially due to cold-load pick-up (CLPU) issues following an extended outage in systems with a high penetration of thermostatically controlled devices \cite{schneider2015evaluating}. Besides, the consecutive switching sequences should not interrupt the service to an already restored distribution customer.

A few have addressed the problem of switching sequence generation for DSR; prominent methods include genetic algorithm, dynamic programming, mixed-integer programming, and sequential optimization methods \cite{watanabe2004genetic, perez2008optimal, thiebaux2013planning, chen2017sequential}.  A majority of the existing algorithms employed to generate the switching sequences work well for conventional distribution systems that do not actively utilize DGs in the restoration process \cite{watanabe2004genetic, perez2008optimal, thiebaux2013planning}. Integrating the operation of DGs into the DSR problem introduces both unprecedented opportunities and challenges when generating a restoration sequence for feeder reconfiguration and active islanding using DGs.  An approach based on particle swarm optimization is proposed for DG allocation for a single-step restoration under CLPU condition \cite{el2012power}. However, this approach is a single step restoration and cannot generate a feasible switching sequence for the restoration plan. To account for the time-critical operations when restoring the distribution grid, an approach based on sequential optimization is proposed in \cite{chen2017sequential}; however, it poses limitations to selecting the length of time-horizon to balance the computation tractability of the algorithm and solution quality.

\subsection{Specific Contributions}
To address the aforementioned challenges, we propose a {\em two-stage service restoration framework} for multi-feeder distribution systems with grid-forming DGs. The proposed DSR framework (1) utilizes all available resources including healthy feeders and grid forming DGs to restore a maximum number of distribution customers and (2) generates a sequence of switching actions that leads to the final restored network configuration via intermediate steps of restored networks that do not violate system operating constraints while considering CLPU and do not disrupt the power supply to an already restored customer. The specific contributions are listed below:
\begin{enumerate}
\item \textit{Two-stage DG-assisted Restoration and Switching Sequence Generation with Col Load Pick-up (CLPU):} 

A two-stage DSR algorithm is proposed that considers feeder reconfiguration and active islanding methods using grid-forming DGs into one unified framework.

\begin{itemize}
    \item [--] The first stage of the problem effectively integrates DG-assisted restoration into traditional DSR problem and enforces active islanding of distribution system, when required, thus ensuring the restoration of a maximum number of distribution customers. It is ensured that the restored network and DG-assisted islands operate in a radial topology via optimal and power-flow feasible restoration paths.

    \item [--] The second stage of the problem generates a sequence of switching actions such that the system can transit from post-faulted condition to final restored system configuration optimally and feasibly. The loss of load diversity is taken into account while considering the behavior of outage loads under CLPU conditions. \end{itemize} 
    
\vspace{0.3cm}    
\item \textit{Scalable MILP formulation for the restoration of a three-phase unbalanced multi-feeder distribution system}: 

Both stages of the proposed methodology are modeled as a mixed-integer linear program (MILP) such that the proposed framework easily scales for a large unbalanced three-phase distribution system. The approach can easily incorporate any new resource (DG/feeder) and/or new switches without significantly increasing the overall computational complexity and generate the sequence of switching operations to optimally restore the distribution systems.
\end{enumerate}

The rest of the paper is organized as the following. Section 2 describes the overall architecture of the proposed DSR framework and its integration within a modern distribution management system (DMS). Sections 3, 4 and 5 introduce the mathematical description and detail formulations for the Stage 1 and Stage 2 of the proposed DSR algorithm. Section 6 presents the simulation results on a 4-feeder 1069-bus distribution test system. Finally, the summary and future research directions are stated in Section 7.

\section{Proposed DSR Framework: Overall Architecture}
A Distribution Management System (DMS) supports grid management and decision support applications for efficient and reliable operation of the electric power distribution systems. FLISR is a distribution automation application that networks groups of switches and available distributed resources on a feeder to vastly improve the reliability of utility delivered power by ``localizing outages'' \cite{vadari2013electric}. 
A FLISR application performs four related actions to restore the power supply after an outage: senses trips in switches that are monitored and controlled by SCADA, locate/identify the faulted sections using triggered protection devices and smart meter pings or customer calls, isolates the faulted section by opening the appropriate switching devices and restores the power supply to the healthy feeders using feeder reconfiguration and intentional islanding using DGs \cite{report1, vadari2013electric}. A distribution service restoration (DSR) algorithm is employed to perform the fourth step that is also most expensive in terms of computational needs. It is worth mentioning that FLISR does not avoid outages but decreases the total outaged time experienced by the distribution customers. 

This article is focused on developing a computationally tractable DSR application for the distribution systems. The proposed DSR application aims to advance the conventional service restoration techniques by coordinating emerging DG technology with other controllable devices (such as switches, breakers) to achieve more efficient restoration solutions. More specifically, the proposed two-stage framework within the DSR will generate a sequence of control actions that optimally coordinate feeder reconfiguration and DG islanding to restore the maximum distribution loads. 

For completeness, the implementation of the proposed DSR method within a DMS is depicted in Fig. \ref{fig:1}.  Once a power outage is detected, the OMS will collect information from the distribution supervisory control and data acquisition (SCADA) system, customer information system (CIS), interactive voice response (IVR), and smart meters, as well as the reports from the field crew. The network model, CLPU load estimation, and resources available for restoration are updated based on the collected information and asset registry in the DMS database. The proposed two-stage DSR algorithm will generate a restoration sequence that can be executed remotely through SCADA or manually by the field crew. The two stages of the proposed DSR algorithm are briefly introduced in this section. The detailed model and formulation for the proposed two-stage DSR framework are described in Sections 4 and 5. 
\vspace{-0.3 cm}
\subsection{Stage 1: Generate Optimal Restoration Topology}
During an outage, protection devices isolate certain parts of a distribution system including the faulty elements. While replacing faulty elements may be time-consuming, existing healthy switches could reconfigure the system to alleviate the outage effect. Thus, in the first stage of the proposed algorithm, an optimal network configuration problem is solved assuming the system transitions instantaneously from the post-fault to the restored configuration. Specifically, given the post-fault status of the network, the first stage of the problem obtains an optimal restored network topology that minimizes the impacts of the outages while complying with the system's operational constraints. The restoration plan includes the possibility of intentional DG-supplied islands (with the grid-forming capability) to restore additional loads. The problem is formulated as an MILP to maximize the restored loads subject to network operational and topological constraints. The decision variables are switch (line/load) statuses and the statuses of grid-forming DGs. 
\vspace{-0.3 cm}
\subsection{Stage 2: Generate Optimal Switching Sequences}
In practice, the transition from post-fault topology to new network topology (with restored loads) requires a sequence of reconfiguration/switching steps involving one control action at a time. Note that the simultaneous action of multiple switches is not recommended as it might lead to temporary meshed/looped topologies that might be detrimental to the distribution system protection scheme designed for radial topologies. Further, due to cold-load pickup (CLPU), a staged restoration is preferred to ensure that the network thermal and voltage limit constraints are not violated. In the second stage of the proposed DSR algorithm, we solve the problem of generating an optimal sequence of switching actions to generate a sequence of power flow feasible restoration plan. The problem objective is to maximize the load restored at each switching step while ensuring that the operational constraints are satisfied. This stage generates the optimal way to transition the faulted system from post-fault condition to restored topology that was generated in Stage 1.

\begin{figure}[t]
    \centering
    \vspace{-0.3cm}
    \includegraphics[width=0.48\textwidth]{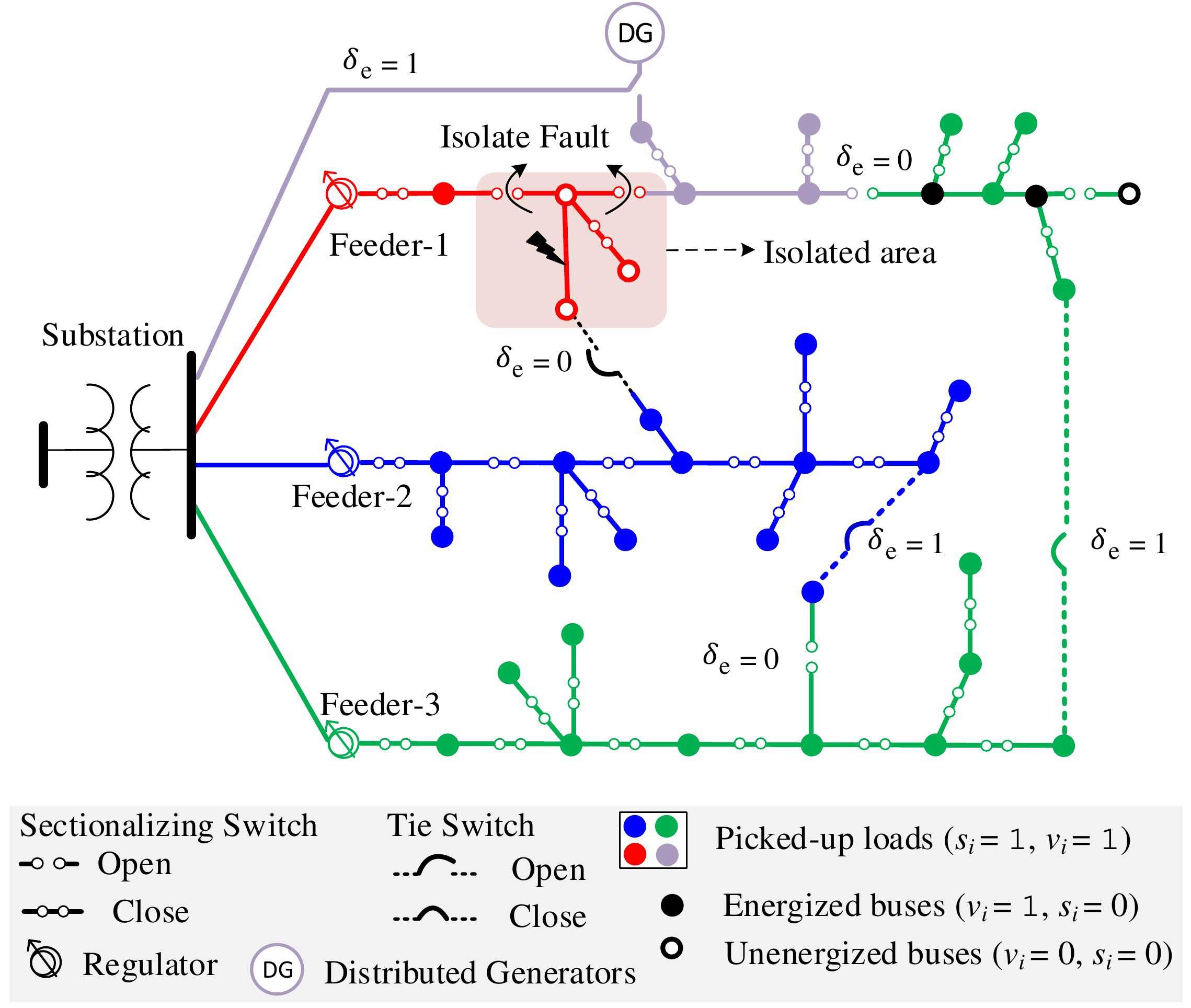}
    \vspace{-0.1cm}
    \caption{An example restoration scenario. The colors indicate energized feeder sections and binary variables define the topology.}
    \label{fig:2}
    \vspace{-0.0cm}
\end{figure}

\section{Grid Modeling, Variables, and Assumptions}
\subsection{Graphical Representation for Distribution System}

We represent the distribution system comprised of multiple feeders and DGs using a connected graph, $\mathcal{G(V, E)}$, where buses are modeled as vertices and physical line sections including switches as edges. Any edge, $e\in \mathcal{E}$, is defined by its incident nodes $(i,j)$ with ($i$, $j$) $\in \mathcal{V}$ where the binary decision variables for buses and edges are shown in Fig. \ref{fig:2}. 

\begin{figure}[t]
    \includegraphics[width=0.48\textwidth]{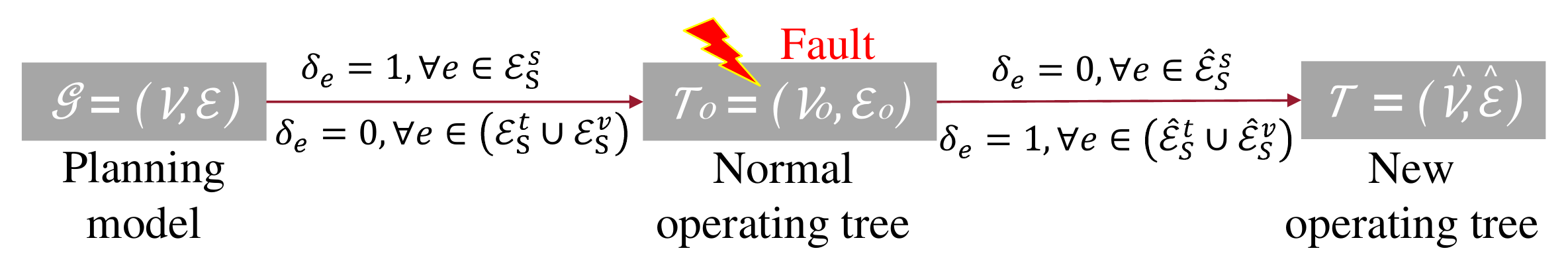}
    \vspace{-0.2cm}
    \caption{Normal operating tree of a planning model and reconfiguration of tree after a fault}
    \label{block}
    \vspace{-0.4cm}
\end{figure}

The normal operating tree of a well-planned distribution network is given as $\mathcal{T}_o = (\mathcal{V}_o, \mathcal{E}_o)$ where all tie-switches are open, all grid-forming DGs are disconnected, and all sectionalizing-switches are closed. Once a fault occurs on a normal operating tree, the proposed DSR algorithm identifies a desired tree/subtrees within the original graph, $\mathcal{G}$, that maximizes the given objective function of restoring loads subject to various connectivity and operating constraints. After the suitable switching scheme is implemented, the new operating tree is defined as $\mathcal{T} = (\hat{\mathcal{V}}, \hat{\mathcal{E}})$ where $\mathcal{\hat{E}}\subseteq \mathcal{E}$ and $\mathcal{\hat{V}}\subseteq \mathcal{V}$ (See Fig. \ref{block}).  The original graph search problem of finding the new operational tree corresponding to the restored distribution system is combinatorial. In this paper, we transform the combinatorial problem into a mathematical programming problem. 

\subsection{Decision Variables}
In this section, we define the binary variables associated with the proposed DSR algorithm. Note that the Stage-1 of the algorithm solves for the binary decision variables defined in this section to obtain an optimal restoration plan. The Stage-2 then generate the sequence in which the decision variables need to be realized to obtain the desired restored distribution system. 

\subsubsection{Bus Energization Variable}
   A binary variable $v_i=\{0,1\}$ is assigned to each bus, where $v_i=1$ implies that bus $i$ is energized, while $v_i=0$ implies bus $i$ is not energized during the restoration process. 

\subsubsection{Load Pick-up Variable}
Each load bus is assigned a binary variable $s_{i}=\{0,1\}$ that represents the switch status of the load connected to the particular bus. This variable helps in the case when only a few critical loads are to be restored without restoring all the loads in the path. Note that for a load to be restored, both $s_i$ and $v_i$ must be 1.

\subsubsection{Switch Variable}
The circuit reconfiguration for restoration requires selecting a subset of switches to be closed and opened to maintain a radial operational topology for the restored network. A binary variable $\{\delta_{e}\}_{e\in \mathcal{E}_S}\in \{0,1\}^{|\mathcal{E}_S|} $ is associated with each switch, where $\delta_{e}=1$ implies that switch connecting buses $i$ and $j$ is closed, while $\delta_{e}=0$ implies that the switch is open. The decision on the line/switch binary variable helps maintain a radial configuration for the restored network. The line variable will be used to formulate power flow constraints and connectivity constraints for the distribution system.

\subsubsection{Grid-forming DG Status Variable}
The complexity of the restoration problem in the distribution system increases significantly in the presence of grid-forming DGs that can intentionally island to restore additional loads. To formulate a unified DSR problem that enables restoration using both DG islands and other feeders, a virtual edge, $\delta_{e}$, is added between the sub-transmission bus and each grid-forming DG as shown in Fig. \ref{fig:2}. The state of this edge determines whether the DG is in isolation mode (OFF) or an island is formed. If the virtual edge is closed, i.e, $\delta_{e}=1$ $(e = i \to j)$, DG located at bus $j$ operates in islanded mode. Note that this DG is virtually connected to sub-transmission bus and therefore can inject three-phase apparent power $\textit{\textbf{P}}_{G}$+j$\textit{\textbf{Q}}_{G}$ to bus $j$. However, to avoid the possible loop configuration, one of the switches in the distribution system needs to be opened. The proposed DSR algorithm takes this constraint into account to enable islanded operation (see Fig. \ref{fig:2}).

\subsection{Assumptions}
The following relevant and justifiable assumptions are made for solving the proposed DSR problem.
\begin{itemize}
    \item[$A_1$:] The required information to solve the DSR problem related to the grid topology and the available resources for restoration are relayed to the DMS. Note that the investment made under the Smart Grid Investment Grant  Program managed by the US  Department of  Energy \cite{SGIGR}  has led to the deployment of remote terminal units,  smart switches, and smart meters in the distribution feeders thus allowing for advanced automation capabilities \cite{sandy, PNNL}. Our assumption is centered on the upgrades made to the distribution systems to allow for implementing the proposed advanced service restoration approach.  \\
    \item [\textbf{$A_2$}:] All grid-forming DGs are equipped with appropriate technology to enable the control of their output voltage and frequency to help realize a stable energized island \cite{rocabert2012control, zarei2018reinforcing}. These grid-forming DGs remain disconnected from the main grid during normal operating conditions. In the case of disruption to the main grid or faults within the distribution system, the DGs are energized to supply additional loads that have been islanded from the main feeders. \\
    \item [$A_3$:] The radial feeder topology is maintained for each restoration path and DGs are not networked. Although DGs can be networked during the restoration process, the operation and control of an islanded distribution network supplied by multiple DGs requires advance control and coordination capabilities that are typically not available in the existing systems \cite{chen2016resilient, gao2016resilience}.\\
    \item [$A_4$:] The fault location and isolation have been implemented before running the proposed service restoration algorithm. Note that the fault location can be identified using OMS based on the information collected from relays, fault indicators, line sensors, and smart meters \cite{report1, chen2017sequential}. The faulted feeder section can be isolated from the rest of the feeder by opening the protection device downstream from the identified fault location. 
\end{itemize}

\section{Stage-1: Find Optimal Restoration Topology}
In this section, we detail the mathematical formulation for the first stage of the restoration problem described in Section II. Specifically, Stage-1 details the MILP formulation to obtain the optimal restoration topology for the given faulted distribution system.

\subsection{Problem Objective}
    The problem objective is to maximize the restored load and minimize the number of switching operations subject to the feeder's operational and connectivity constraints.
    \subsubsection{Maximize the Restored Load}
    The first objective is to maximize the amount of load restored while considering different weight factors for each load ($w_i$) that indicate load priority. 
    The objective is defined as the following.
    \begin{equation}
    \text{Maximize   } \sum_{i\in  \mathcal{V}_S}\sum_{\phi\in\{a,b,c\}}w_i\ s_i\ P_{Li}^{\phi}.
    \end{equation}
    \subsubsection{Minimize the Number of Switching Operations}
     The number of switching operation determines the performance of the restoration plan as it closely relates to the time taken to execute the restoration plan. Also, frequently operating switches adds additional maintenance cost. Therefore, it is desirable to minimize the number of switching operations so that the restoration plan can be executed in an efficient and timely manner. Thus, the second objective is to minimize the total number of switching operations defined in (\ref{eq:sec}).
    \begin{equation}\label{eq:sec}
    \text{Minimize   } \Big(\sum_{e\in \mathcal{E}_S^s}(1-\delta_{e})+\sum_{e\in \mathcal{E}_S^t}\delta_{e} +\sum\limits_{e \in  \mathcal{E}_S^v} \delta_{e}   \Big).
     \end{equation}
\subsubsection{Weighted Objective Function}
We define a multi-objective restoration problem using a weighted combination of the two previously defined objective functions in (\ref{eq0}). 
      
As the customer's satisfaction depends upon the availability of the power supply, the maximization of the restored load is defined as the primary objective and is always given a higher preference. The minimization of the total number of switching operations is defined as the secondary objective. The weights, $\alpha$, $\beta$, and $\gamma$ are defined such that the primary objective is always prioritized. Since the secondary objective is a sum of binary variables only, making $\beta <1, \gamma <1$, and assigning $\alpha$ a large number ensures that the problem first restores the maximum weighted loads and then minimizes the switching operations. Also, DGs must not operate in an islanded mode unless it is required. 
Thus, the priority for restoration is to use feeder backup by switching a pair of sectionalizing and tie switches. When a normally open tie switch operates, a normally closed sectionalizing switch has to be opened to maintain a radial topology. Thus, gamma is made at least $2|\mathcal{E}_S^t|$ times higher than $\beta$ (i.e, $\gamma \geq 2|\mathcal{E}_S^t|\beta$).

\subsection{Problem Constraints}
The several constraints associated with the proposed service restoration problem are described in this section.
\subsubsection{Connectivity Constraints} This section defines the set of topological constraints required to ensure a connected network for the restored loads and a radial restored network/s topology.
\begin{figure}[!t]
\begin{minipage}[t]{0.48\textwidth}
\noindent\rule{8.6cm}{1.2pt}\\
\textbf{Stage 1: Final Optimal Configuration}\\
\noindent\rule{8.6cm}{1.2pt}\\
\textbf{Maximize}:
    \begin{eqnarray} \label{eq0}
        \small
        \begin{aligned}
        & \Bigg(\alpha \sum_{i\in  \mathcal{V}_S}\sum_{\phi\in\{a,b,c\}}w_i\ s_i \ P_{Li}^{\phi}-\beta \Big(\sum_{e\in \mathcal{E}_S^s}(1-\delta_{e})\\
        &-\sum_{e\in \mathcal{E}_S^t}\delta_{e}\Big) -\gamma\sum\limits_{e \in  \mathcal{E}_{S}^v} \delta_{e}   \Bigg).
        \end{aligned}
    \end{eqnarray}
\noindent\rule{8.6cm}{.3pt}\\
\textbf{Subject to}: \vspace{-0.5 cm}
   \begin{subequations}\label{connect}
        \begin{align}
        s_i&\leq v_i, \ \  \forall i\in \mathcal{V}_S \\
        s_i &=v_i, \ \  \forall i\in \mathcal{V}\textbackslash \mathcal{V}_S.
        \end{align}
    \end{subequations}
    \vspace{-0.8 cm}
    \begin{subequations}\label{lines}
        \small
        \begin{align}
        \delta_{e}&\leq v_i,\  \delta_{e}\leq v_j, \  \forall e\in \mathcal{E}_S\textbackslash\mathcal{E}_F \\
        \delta_{e}&= 0, \ \  \forall e\in \mathcal{E}_F.
        \end{align}
    \end{subequations}
     \begin{equation}\label{radial}
     \small
        \sum_{e\in \mathcal{E}_c} \delta_{e}\leq |\mathcal{E}_c|-1, \hspace{0.6 cm}  \forall e\in \mathcal{E}_c.
    \end{equation}
    \vspace{-0.5 cm}
    \begin{subequations}\label{pflow}
    \small
    \begin{align}
     \sum_{e:(i,j) \in \mathcal{E}} \textbf{\textit{P}}_{e} &= s_j\ \textbf{\textit{P}}_{Lj}+ \sum_{e:(j,i) \in \mathcal{E}}                                 \textbf{\textit{P}}_{e} \\
        \sum_{e:(i,j) \in \mathcal{E}}\textbf{\textit{Q}}_{e} &= s_j\ \textbf{\textit{Q}}_{Lj}+ \sum_{e:(j,i) \in \mathcal{E}}                 \textbf{\textit{Q}}_{e}\\
    \textbf{\textit{U}}_{i}-\textbf{\textit{U}}_{j} &= 2\big( \tilde{\textbf{r}}_{e} \textbf{\textit{P}}_{e}+\tilde{\textbf{x}}_{e} \textbf{\textit{Q}}_{e}),\ \  \forall e\in \mathcal{E}\textbackslash(\mathcal{E}_S\cup \mathcal{E}_R)\\
    \delta_{e}\ (\textbf{\textit{U}}_{i}-\textbf{\textit{U}}_{j}) &= 2\big( \tilde{\textbf{r}}_{e} \textbf{\textit{P}}_{e}+\tilde{\textbf{x}}_{e} \textbf{\textit{Q}}_{e}),\ \  \forall e\in \mathcal{E}_S.
        \end{align}
    \end{subequations}
    \vspace{-0.8 cm}
    \begin{subequations}\label{vr}
    \begin{align}
         V_j^\phi &= a_\phi  V_i^\phi,\\
          \textbf{\textit{U}}_j &= A^\phi \textbf{\textit{U}}_i,\ \  \ \forall e:(i,j)\in \mathcal{E}_R. 
    \end{align}
    \end{subequations}
     \begin{equation} \label{eq2}
        \small
        q_{cap,i}^{\phi} = u_{cap,i}^\phi q_{cap,i}^{rated,\phi} U_{i}^\phi.
    \end{equation}
    
    \begin{equation}\label{voltlim}
        \small
        v_i  \textbf{\textit{U}}^{min}\leq \textbf{\textit{U}}_{i}\leq v_i  \textbf{\textit{U}}^{max}, \ \ \forall i\in \mathcal{V}.
    \end{equation}
    \begin{equation}\label{flowlim}
       \small
       \left(\textbf{\textit{P}}_{e}\right)^{2} + \left(\textbf{\textit{Q}}_{e}\right)^{2} \leq  \left(\textbf{\textit{S}}_{e}^{rated}\right)^{2} \ \ \forall e\in\mathcal{E}\textbackslash \mathcal{E}_S.
    \end{equation}
    \begin{small}
        \begin{equation}\label{quad}
           \begin{aligned}
           -\sqrt{3}\ (\textbf{\textit{P}}_{e}+\textbf{\textit{S}}_{e})&\leq \textbf{\textit{Q}}_{e}\leq -\sqrt{3}\ (\textbf{\textit{P}}_{e}-\textbf{\textit{S}}_{e}), \\
           -\sqrt{3}/2\ \textbf{\textit{S}}_{e}&\leq \textbf{\textit{Q}}_{e}\leq \sqrt{3}/2\ \textbf{\textit{S}}_{e},\\
           \sqrt{3}\ (\textbf{\textit{P}}_{e}-\textbf{\textit{S}}_{e})&\leq \textbf{\textit{Q}}_{e}\leq \sqrt{3}\ (\textbf{\textit{P}}_{e}+\textbf{\textit{S}}_{e}),\ \ \forall e\in \mathcal{E}\textbackslash\mathcal{E}_S.
       \end{aligned}
    \end{equation}
    \end{small}
    \begin{equation}\label{linelim}
    \begin{small}
           \delta_{e}\big[\underline{M}_p \ \ \underline{M}_q\big]\leq \big[\textbf{\textit{P}}_{e}\ \ \textbf{\textit{Q}}_{e}\big] \leq \delta_{e}\big[\bar{M}_p \ \ \bar{M}_q\big], \ \ \forall e\in\mathcal{E}_S.
    \end{small}
    \end{equation}
    \begin{equation}\label{dglim}
    \begin{small}
           \sum_{\phi\in \{a,b,c\}}^{}P_{e}^\phi\leq \delta_{e} P_{G}^{max} \hspace{0.1cm} \text{,} \hspace{0.1cm}
           \sum_{\phi\in \{a,b,c\}}^{}Q_{e}^\phi\leq  \delta_{e} Q_{G}^{max}, \ \  \forall e\in\mathcal{E}_S^v.
    \end{small}
    \end{equation}

\noindent\rule{8.6cm}{.3pt}\\
\textbf{where}: \\
$\tilde{\textbf{r}}_{e}= \text{Real}\{\alpha \alpha^H\} \otimes \textbf{r}_{e}+\text{Im}\{\alpha \alpha^H\} \otimes \textbf{x}_{e}$,$\tilde{\textbf{x}}_{e}= \text{Real}\{\alpha \alpha^H\} \otimes \textbf{x}_{e}+\text{Im}\{\alpha \alpha^H\} \otimes \textbf{r}_{e}$, $\alpha=[1\ \  e^{-j2\pi/3} \ \ e^{j2\pi/3}]^T$ \\ \\
$\textbf{\textit{S}}_{e}=\textbf{\textit{S}}_{e}^{rated}\sqrt{(2\pi/n)/\text{sin}(2\pi/n)}$ and $n=6$\\
$a_\phi = \sum\limits_{i=1}^{32} b_i u_{tap,i}^\phi$ and  $\sum\limits_{i=1}^{32} u_{tap,i}^\phi = 1$\\
\noindent\rule{8.6cm}{1.2pt}\\
\end{minipage}
\vspace{-0.9 cm}
\end{figure}
\begin{itemize} 
\item Constraint (\ref{connect}a) ascertains that a load with a switch can be picked up if and only if it is connected to a bus that is energized in the restored network by one of the feeders or DGs. Constraint (\ref{connect}b) ensures that a non-switchable load will be energized depending upon the associated bus energization variable. Thus, a non-switchable load is always picked up if the corresponding bus is energized.\\
\item The constraints for decision upon line energization variable, ($\delta_{e}$), are defined in (\ref{lines}a-\ref{lines}b). The set of equations implies that the decision on lines to be used in the restoration depends upon the corresponding buses and their energization statuses. Equation (\ref{lines}a) ensures that if a line with a switch is energized, the buses connecting the line must be energized. The faults and the open switches in the distribution network are modeled using constraint (\ref{lines}b). \\
\item A new radial configuration of the faulted distribution system is achieved by closing and opening the appropriate tie switches, sectionalizing switches, and virtual edges. A radial topology for restored network/s is ensured using constraint (\ref{radial}) that enforces at least one of the switches in any cycle to be open. All possible cycles in a distribution network are enumerated using iterative loop counting algorithm which is a ``brute force'' technique \cite{ilca}. Then, (\ref{radial}) is written for each cycle. The number of cycles in a graph increases with the increase in the number of tie switches. Despite that, $\mathpzc{G}$ is usually sparse for a distribution network where the total number of cycles is much less than $2^{|\mathpzc{V}|}$. It is important to note that this enumeration can be done offline by storing all the cycles as it is fixed for a given planning model of a distribution network. Thus, the enumeration approach does not affect the real-time computational complexity.
\end{itemize}

\subsubsection{Power Flow Constraints}
A three-phase linearized AC power flow model for the unbalanced distribution system proposed in \cite{gan2014convex} is used in this work. The linearized model is sufficiently accurate and applicable for restoration problems \cite{gan2014convex}. The restoration problem requires the decision upon which lines are energized while accounting for network operating constraints. The power flow along a line is only valid if the line is energized. Therefore, to appropriately represent the restoration problem the branch flow equations are coupled with line and bus decision variables.

\begin{itemize} 
\item  Constraints (\ref{pflow}a-\ref{pflow}d) represent three-phase unbalanced linearized power flow equations and relates power injections and voltages. After ignoring the power losses, constraints (\ref{pflow}a) and (\ref{pflow}b) define active and reactive power flow for each lines $e \in \mathcal{E}$. Similarly, constraint (\ref{pflow}c) defines voltage equations for non-switchable line while for switchable lines, the voltage drop applies only if the switch is closed (\ref{pflow}d). Note that constraint (\ref{pflow}d) is non-convex as it involves product of variables. These constraints are linearized by defining an auxiliary variable and using the big-M method \cite{winston2003introduction}. 
The big-M method replaces the product of variables by their linear convex envelopes to yield a relaxation of the original non-convex feasible set. Take for instance the product $z = xy$
over a binary variable $x\in\{0,1\}$, and a continuous variable
$y$ bounded within $y \in [\underline{y}, \bar{y}]$. The constraint $z = xy$ can be
equivalently expressed as four linear equality constraints.
\begin{subequations}\label{bigm}
    \begin{align}
    x\underline{y} &\leq z\leq x\bar{y},\\
    y+(x-1)\bar{y} &\leq z\leq y+(x-1)\underline{y}.
    \end{align}
\end{subequations}
To see the equivalence, substitute $x = 0$ in (\ref{bigm}) to get $z = 0$, and put $x = 1$ to get $z = y$. Note that all continuous-binary bilinear products encountered henceforth will be handled using the big-M method.
\end{itemize}

\subsubsection{Voltage Regulator and Capacitor Banks Models}
 For a large feeder, voltage regulators (VR) and capacitor banks (CB) may help restore additional loads that might not be possible due to potential undervoltage issues. In this section, we detail the mathematical model for VR and CB where VR tap position and CB status are modeled as binary variables. 

A 32-step VR with a voltage regulation range of $\pm10\%$ is assumed. Let, $a_\phi$ be the turn ratio for the VR connected to phase $\phi$ of line $e:(i,j)\in \mathcal{E}_R$. Then $a_\phi$ can take values between 0.9 to 1.1 with each step resulting in a change of 0.00625 pu. 
Let, for $u_{tap,i}^\phi \in \{0,1\}$ be a binary variable defined for each regulator step position i.e. $i \in (1,2,...,32)$. Define a vector $b_i \in \{0.9, 0.90625 , ..., 1.1\}$. Then, (\ref{vr}a) gives the voltage relation between nodes $i$ and $j$. Taking square of voltage equation, defining $a_\phi^2 = A_\phi$, $b_i^2 = B_i$, and realizing that $(u_{tap,i}^\phi)^2 = u_{tap,i}^\phi$ we obtain (\ref{vr}b) representing model for VR. Note that if a multi-phase VR is gang operated, a single $A^\phi$, and $u_{tap,i}^\phi$ variables are defined for each phase.

The reactive power generated by CB, $q_{cap,i}^{\phi}$, is defined as a function of binary control variable $u_{cap,i}^\phi \in \{0,1\}$ indicating the status (On/Off) of the CB, its rated per-phase reactive power $q_{cap,i}^{rated,\phi}$, and the square of the bus voltage at bus $i$ for phase $\phi$, $U_{i}^\phi$ in (\ref{eq2}). The CB model is assumed to be voltage dependent and provides reactive power as a function of $U_{i}^\phi$ when connected, i.e. $u_{cap,i}=1$. For a three-phase CB, a common control variable, $u_{cap,i}^\phi$, is defined for each phase.
Both (\ref{vr}) and (\ref{eq2}) include a product of binary and continuous variables that can be easily linearized using the big-M method.  

\subsubsection{Network Operating Constraints} This section defines nodal voltage limit constraints and thermal limit constraints for lines and transformers.
\begin{itemize} 
\item The voltage of each bus should be within the limit as specified in ANSI C84.1 and is ensured by equation (\ref{voltlim}). $\textbf{\textit{U}}^{min}$ and $\textbf{\textit{U}}^{max}$ are set to $(0.95)^2$ and $(1.05)^2$, respectively for each phase of the bus.\\
\item The loading on the lines and transformers must not exceed the rated kVA capacity. The rated kVA capacity is specified for the transformers. The thermal limit for a line is, however, specified in terms of its ampacity. We use a nominal feeder voltage of 1 p.u. to convert line ampacity rating to their rated kVA capacity. The actual thermal limit constraint is specified using the quadratic equation in (\ref{flowlim}). We use the polygon-based linearization approach proposed in \cite{ahmadi2016linear} to linearize (\ref{flowlim}) by a set of linear constraints defined in (\ref{quad}). 
We use (\ref{quad}) instead of (\ref{flowlim}) in the MILP model.\\
\item For reconfiguration purpose, it is required to force the line flow in open switches to be zero. If switch connecting buses $i$ and $j$ is open ($\delta_{e} = 0$), constraint (\ref{linelim}) sets the power flow on the line $(i,j)$ to 0. Else, box constraints on the power flow are enforced where $\underline{M}_p = - \bar{M}_p$ and $\underline{M}_q = -\bar{M}_q$.\\

\end{itemize}

\vspace{-0.3 cm}
\subsubsection{DG Operating Constraints} Constraint (\ref{dglim}) ensures that the in-flow power of each DG from the sub-transmission bus should be less than or equal to the rated DG capacity. This is per the fact that all DGs are connected to the sub-transmission bus using virtual edges to assist with intentional islanding via grid forming DGs.

\begin{figure}[t]
\centering
    \includegraphics[width=0.4\textwidth]{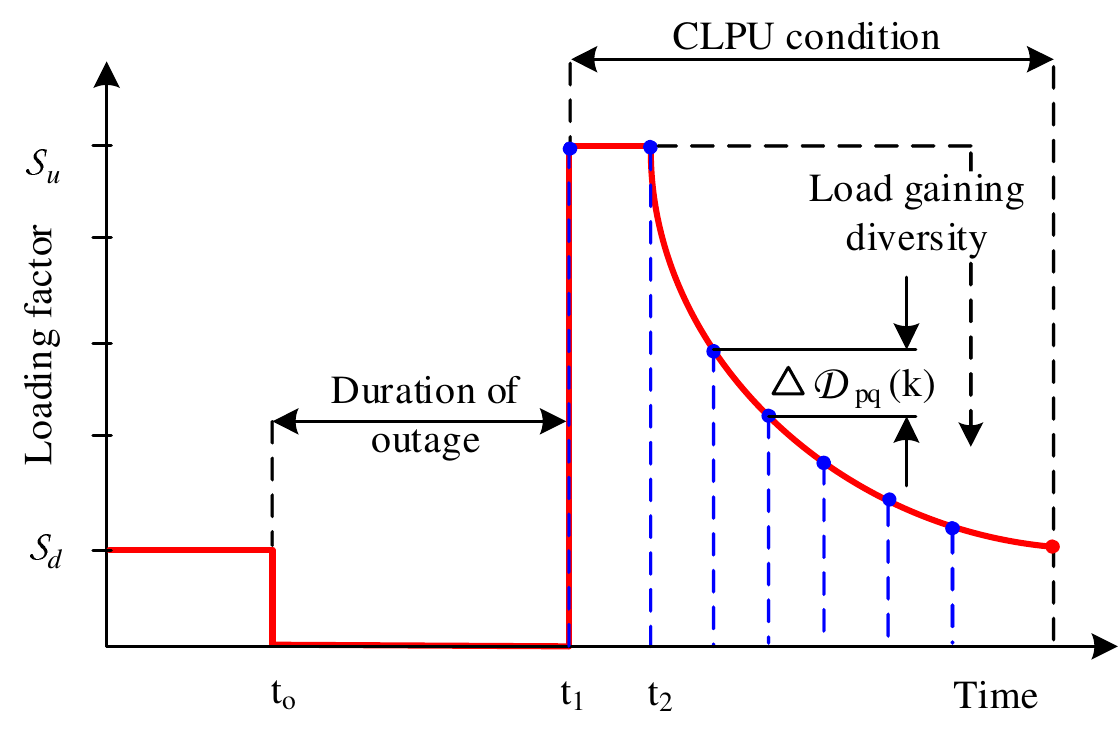}
   \vspace{-0.2cm}
    \caption{CLPU event for feeder demand }
    \label{clpu}
    \vspace{-0.0cm}
\end{figure}

 \begin{figure}[!t]
\begin{minipage}[t]{0.47\textwidth}
\noindent\rule{8.57cm}{1.2pt}\\
\textbf{Stage 2: Optimal Switching Scheme}\\
\noindent\rule{8.57cm}{1.2pt}\\
\textbf{Maximize}:
\begin{equation}
  \sum_{t \in  {T}} \sum_{i \in   \mathcal{V}_S}\sum_{\phi\in\{a,b,c\}}w_is_{i,t}P_{Li,t}^{\phi}
\end{equation}
\noindent\rule{8.57cm}{.2pt}\\
\textbf{Subject to}:\\ 
${T}$ copies of Stage 1 constraints, variables indexed by $t$.
\vspace{-0.0 cm}
\begin{subequations}\label{stage2}
\small
    \begin{align}
        &s_{i,t} \geq s_{i,t-1} \hspace{0.4 cm} \forall i\in \mathcal{V}_S, \forall t \in {T} \\
        &s_{i,t} = 1 \hspace{0.4 cm} \forall i\in \hat{\mathcal{V}}_S, \  t = {T_{|T|}}\\ 
        &\delta_{e,t-1}  \geq \delta_{e,t} \hspace{0.4 cm} \forall e \in \hat{\mathcal{E}}_S^s, \forall t \in {T}\\ 
        &\delta_{e,t}  \geq \delta_{e,t-1} \hspace{0.4 cm} \forall e \in \big(\hat{\mathcal{E}}_S^t \cup \hat{\mathcal{E}}_S^v\big), \forall t \in {T}\\ 
        &\sum_{e \in  \hat{\mathcal{E}}_S^s}  \big(\delta_{e,t-1} - \delta_{e,t}\big) + \sum_{e \in  (\hat{\mathcal{E}}_S^t\cup \hat{\mathcal{E}}_S^v)} \big(\delta_{e,t} - \delta_{e,t-1}\big) <= 1,  \hspace{0.1 cm} \forall t \in {T}\\ 
        &\delta_{e,t} = 0 \hspace{0.4 cm} \forall e\in \hat{\mathcal{E}}_S^s, \  t = {T_{|T|}}\\
        &\delta_{e,t} = 1 \hspace{0.4 cm} \forall e\in \big(\hat{\mathcal{E}}_S^t \cup \hat{\mathcal{E}}_S^v\big), \  t = {T_{|T|}}\\
         &\delta_{e,t} = 0 \hspace{0.4 cm} \forall e \in \{\mathcal{E}_S^t - \hat{\mathcal{E}}_S^t\}, \  t \in {T}\\
         &\delta_{e,t} = 0 \hspace{0.4 cm} \forall e \in \{\mathcal{E}_S^v - \hat{\mathcal{E}}_S^v\}, \  t \in {T}\\
         &\delta_{e,t} = 1 \hspace{0.4 cm} \forall e \in \{\mathcal{E}_S^s-\hat{\mathcal{E}}_S^s\}, \  t \in {T}.
        \vspace{-0.3 cm}
   \end{align}
\end{subequations}
\noindent\rule{8.57cm}{.2pt}\\
\textbf{where}:\\
    $\hat{\mathcal{E}}_S^s \subseteq \mathcal{E}_S^s$, $\hat{\mathcal{E}}_S^t \subseteq \mathcal{E}_S^t$, and $\hat{\mathcal{E}}_S^v \subseteq \mathcal{E}_S^v$  are the set of sectionalizing switches, tie switches, and DG switches respectively to operate to reach the final configuration. Similarly, $\hat{\mathcal{V}}_S$ is the set of nodes to be energized by the final time step as per solution from Stage 1.
\noindent\rule{8.57cm}{1.2pt}\\
\end{minipage}
\vspace{-0.5 cm}
\end{figure}

\begin{figure*}[t]
    \centering
    \includegraphics[width=0.89\textwidth]{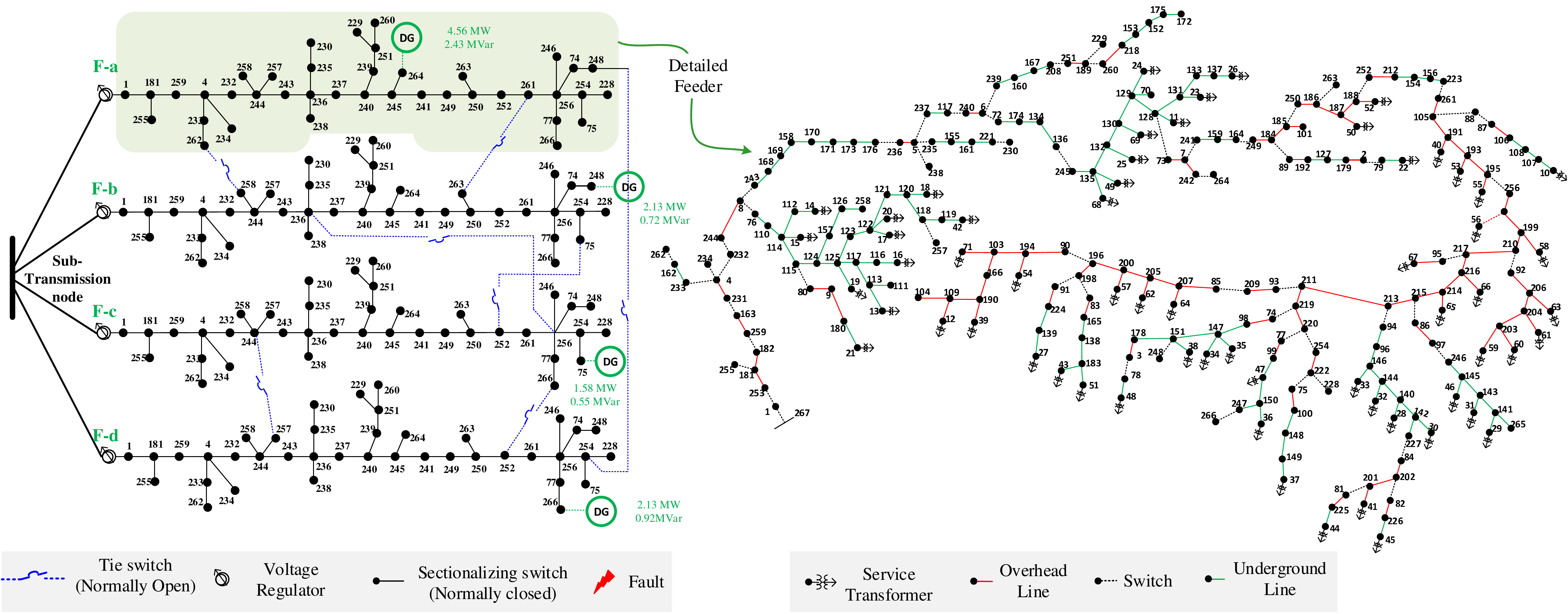}
   \vspace{-0.1cm}
    \caption{Simplified one-line diagram of the multi-feeder 1069-bus distribution system with seven additional tie switches and four DGs.}
    \label{test_case}
   \vspace{-0.1 cm}
\end{figure*}

\section{Stage-2: Find Optimal Switching Sequence}
 
In this section, we detail the problem formulation for the Stage-2 of the proposed restoration algorithm. The Stage-2 of the problem generates an optimal switching sequence to help realize the transition of the faulted system to the optimal restoration topology obtained in the Stage-1. The formulation includes the effects of cold load pick up (CPLU) constraints in generating the switching sequences. The problem objective is to obtain the sequence in which the sectionalizing switches, tie switches, and load switches are to operate to enable the transition of the distribution system from $\mathcal{T}_o$ to $\mathcal{T}$ without violating system's operating constraints during the intermediate steps. In the following section, first, we describe the load modeling for the CPLU event. Next, we detail the MILP model to generate the optimal switching sequence. 

\subsection{Load Modeling during CLPU Event}
CLPU is observed when restoring the service to end-use loads specifically those with a regulated process such as irrigation pumps or thermostatically controlled devices. In power distribution systems, due to the high penetration of thermostatically and/or process controlled end-use loads, the distribution system can experience a significantly higher demand when restoring the out-of-service area for outages as short as 10 minutes. Out of the several causes for CLPU, loss of load diversity among the thermostatically controlled
loads, which can persist for tens of minutes to hours is of greater concern as it increases the thermal loading of equipment and thus restricting the restoration of network due to the violations in distribution system's operating constraints \cite{friend2009cold, el2012power}. In existing literature, several models have been proposed to estimate the behavior of the loads under CLPU conditions including physical models, exponent models, probabilistic models, and regression-based models \cite{schneider2015evaluating,leou1996distribution}. However, the resultant CLPU characteristics are found to be in close agreement with a delayed exponential model \cite{lang1982analytical}, expressed mathematically in (\ref{clpu1}).
\vspace{-0.3cm}
\begin{eqnarray} \label{clpu1}
\begin{aligned}
 \mathpzc{D}_{pq}(k) =& \mathpzc{S}_d + (\mathpzc{S}_u - \mathpzc{S}_d). e^{-\alpha(k-t_i)}\\
  + & \mathpzc{S}_u[1-u(k-t_i)].u(k-{T}_i).
  \vspace{-0.3 cm}
\end{aligned}
\end{eqnarray}

\noindent where, $\mathpzc{D}_{pq}(k)$ is the scale factor on the CLPU curve at $k^{th}$ sample,  $\mathpzc{S}_u$ is the undiversified loading factor, $\mathpzc{S}_d$ is the diversified loading factor, $\alpha$ is the rate of load decay, ${T}_i$ is the restoration time ($t_2$), and $u(.)$ is a step function. One of the challenges with the formulation detailed in (\ref{clpu1}) is determining the correct delay time. This is especially challenging since these values vary with time of day, season, and length of the outage. One of the potential solutions is to use a stochastic modeling approach. Note that estimating the parameters of the CLPU model described in (18) is beyond the scope of this paper. Please refer to \cite{mortensen1988stochastic, athow1994development} for further information on CLPU models. 

In this paper, we employ the delayed exponent CLPU model as expressed in (\ref{clpu1}). We perform a piece-wise linearization of the given delayed exponent curve modeling the CLPU characteristics. The piece-wise linearization is done to avoid the non-linearity in the resulting switching sequence generation problem. Fig. \ref{clpu} shows a delayed exponent model for a load. Note that in this model, the peak CLPU current is maintained for a specified amount of time following which the load demand decays exponentially. The outage occurs at $t_0$, and the load is energized at $t_1$. Due to the loss in load diversity, the undiversified loading factor at $t_1$ is $\mathpzc{S}_u$. The load starts to gain diversity at $t_2$ and decreases exponentially. The post-outage diversified loading factor is normally equal to the pre-outage loading level $\mathpzc{S}_d$. Note that the restoration time ($t_1$) for each load is not pre-determined. However, from the given CLPU curve, we can calculate the load demand at each sampling time. For simplicity, the CLPU curve is equally sampled and the sampled values are picked up for the load at consecutive time steps once the load energizing variable $s_i$ becomes equal to  1. Assuming we discretize the CLPU curve for a particular load into total $N$ equally sampled steps, the difference between two scale factors at $k^{th}$ sample and $(k-1)^{th}$ sample is given by (\ref{eq:dif}). Note that $\Delta \mathpzc{D}_{pq}(k)$ can be replaced with $ \Delta P_{Li}(k)$ and $\Delta Q_{Li}(k)$ for formulating $\textbf{\textit{P}}_{Li,t}$ and $\textbf{\textit{Q}}_{Li,t}$ respectively (See equation(\ref{pqclpu})).
\vspace{-0.3 cm}
\begin{equation}\label{eq:dif}
   \Delta \mathpzc{D}_{pq}(k)  =  
    \begin{cases}
    0,& k = 1\\
   \mathpzc{D}_{pq}(k)-\mathpzc{D}_{pq}(k-1),& 1 < k \leq N \\
    \end{cases}
\end{equation}
\vspace{-0.3 cm}
\begin{subequations}\label{pqclpu}
\small
    \begin{align}
        \textbf{\textit{P}}_{Li,t} &= \textbf{\textit{P}}_{Li}\Big(\mathpzc{S}_u \ s_{i,t} - \sum_{k = 1}^{N} \Delta P_{Li}(k) \  s_{i,t-k+1} \Big) \\
         \textbf{\textit{Q}}_{Li,t} &= \textbf{\textit{Q}}_{Li}\Big(\mathpzc{S}_u \ s_{i,t} - \sum_{k = 1}^{N} \Delta Q_{Li}(k) \  s_{i,t-k+1} \Big).
    \end{align}
\end{subequations}
where, $\Delta P_{Li}(k)$ and $\Delta Q_{Li}(k)$ are calculated using (\ref{clpu1}) and (\ref{eq:dif}). Once CLPU demand is obtained for a particular time-step, $t$, i.e., $\textbf{P}_{Li,t}$ and $\textbf{Q}_{Li,t}$, the outaged load in (\ref{pflow}) is replaced by the equivalent CLPU load demand for the Stage-2 problem. Note that to include CLPU effect in restoration problem, we coordinate the load control switch ($s_i$) along with the sectionalizing and tie-switches in the Stage-2 problem.

\begin{table*}[t]
    \centering
    \small
    \caption{ Restoration strategy for the four-feeder 1069-bus test case without DGs.}
    \label{case1}
    \vspace{0.2 cm}
    \begin{tabular}{c|c|c|c|c|c|c|c}
        \hline
            \multirow{2}{*}{Scenario} & \multirow{2}{*}{Line Fault} & Switch&Load Loss  &\multirow{2}{*}{ Isolation} & \multicolumn{2}{c|}{Switching Schemes without DGs}& Load Restored \\
            & & Tripped & kW && Open & Close &kW\\
            \hline
            \multirow{4}{*}{1} & \multirow{4}{*}{241-159 (F-d)} & \multirow{4}{*}{7-241 (F-d)} & \multirow{4}{*}{3627.56}& \multirow{4}{*}{164-249 (F-d)}& \multirow{4}{6.7 em}{220-254 (F-d)$\tiny{^{1}}$ 136-245 (F-b)$\tiny{^{3}}$ 195-256 (F-c)$\tiny{^{5}}$ }& 248-254 (F-a and F-d)$\tiny{^{7}}$ &\multirow{4}{*}{2973.5}\\
            &&&&&&266-252 (F-c and F-d)$\tiny{^{2}}$&\\
            &&&&&&75-252 (F-b and F-c)$\tiny{^{4}}$ &\\
            &&&&&&236-256  (F-b and F-c)$\tiny{^{6}}$&\\
            \hline
            
            \multirow{6}{*}{2} & \multirow{6}{*}{181-182 (F-c)} & \multirow{6}{*}{267-1 (F-c)} &\multirow{6}{*}{4366.95}& \multirow{6}{*}{182-259 (F-c)}& \multirow{6}{6.7 em}{77-220 (F-c)$\tiny{^{2}}$ 177-240 (F-b)$\tiny{^{8}}$ 195-256 (F-b)$\tiny{^{6}}$ 208-251 (F-c)$\tiny{^{4}}$ 220-254 (F-d)$\tiny{^{10}}$}& 261-263 (F-a and F-b)$\tiny{^{9}}$ &\multirow{6}{*}{3985.20}\\
            &&&&&& 248-254 (F-a and F-d)$\tiny{^{11}}$&\\
            &&&&&& 244-257 (F-c and F-d)$\tiny{^{3}}$ &\\
            &&&&& &236-256 (F-b and F-c)$\tiny{^{5}}$&\\
            &&&&& &252-75 (F-c and F-b)$\tiny{^{7}}$&\\
            &&&&& &252-266 (F-d and F-c)$\tiny{^{1}}$&\\
        \hline
        
        \multirow{3}{*}{3} & \multirow{3}{5.99 em}{193-195 (F-b) 105-191 (F-d) } & \multirow{3}{5.99 em}{223-261 (F-b) 223-261 (F-d) } & \multirow{3}{*}{6757.82}& \multirow{3}{5.7 em}{195-256 (F-b) 195-256 (F-d)}& \multirow{3}{6.7 em}{195-256 (F-c)$\tiny{^{2}}$} & 248-254 (F-a and F-d)$\tiny{^{1}}$&\multirow{3}{*}{4855.9}\\
        &&&&&&75-252 (F-b and F-c)$\tiny{^{4}}$&\\
        &&&&&&266-252 (F-c and F-d)$\tiny{^{3}}$&\\
        \hline
    \end{tabular}
    \vspace{-0.3 cm}
\end{table*}

\begin{figure*}[t]
\centering
\begin{subfigure}{.49\textwidth}
  \centering
  \includegraphics[width=.99\linewidth]{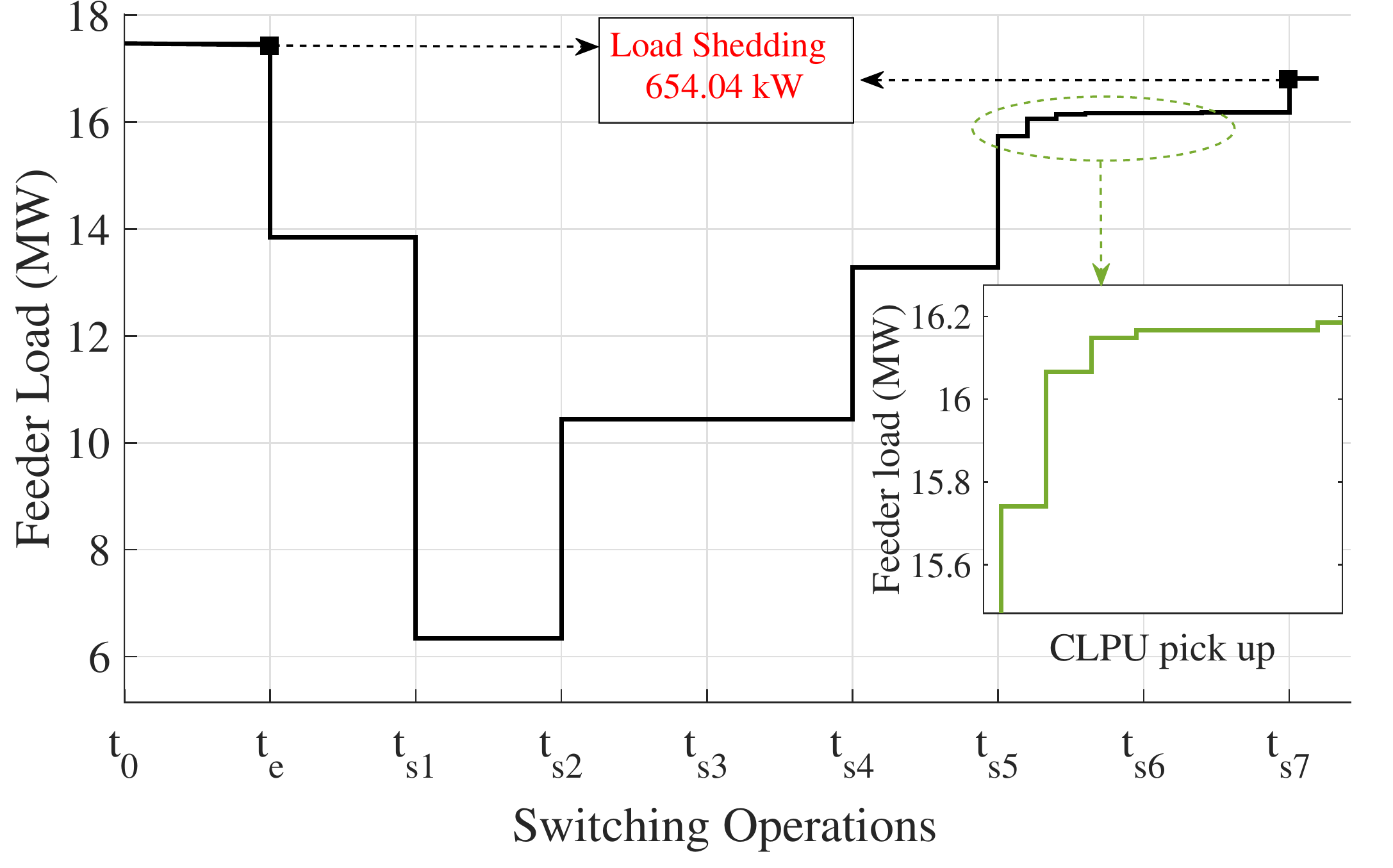}
  \caption{\small Feeder load picked up after sequence of switching operations. CLPU effect is highlighted in green color}
  \label{fig:sub2_1}
\end{subfigure}
\begin{subfigure}{.49\textwidth}
  \centering
  \includegraphics[width=.9\linewidth]{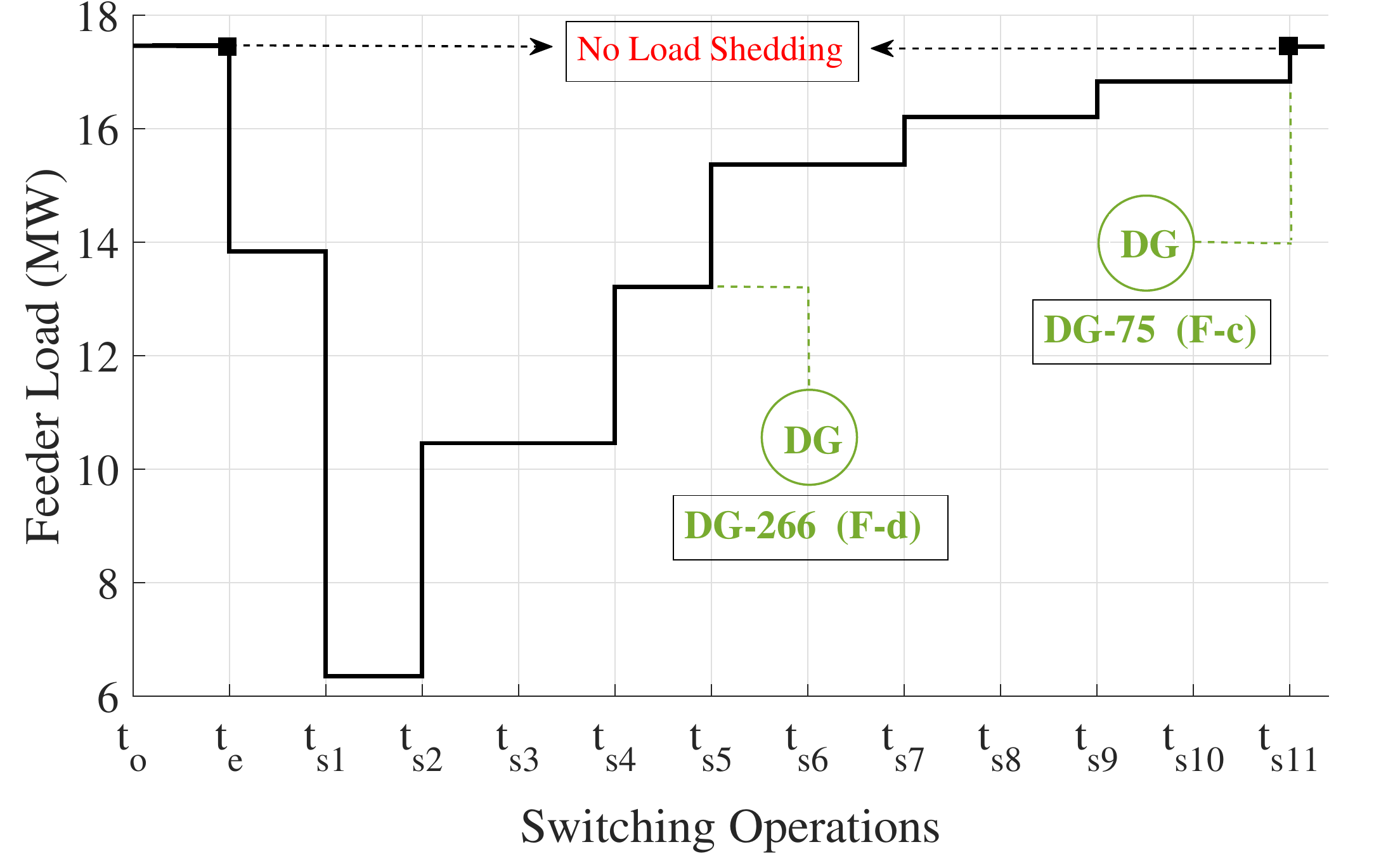}
  \caption{\small Feeder load picked up after sequence of switching operations including DG islanding. CLPU effect is neutralized with DG operation}
  \label{fig:sub2_2}
\end{subfigure}
\caption{\small Switching sequence for Scenario 1 (Fault at Line 241-159 (F-d))}
\label{fig:sc1}
\vspace{-0.3 cm}
\end{figure*}

\subsection{Optimal Switching Sequence}

The restoration of the loads requires toggling of the switches/devices whose initial and final positions differ as per the solution from the Stage-1. This can easily be done without violating the operational constraints simply by performing all of the opening actions before the closings. This naive ordering may interrupt additional customers during the restoration process that need not be outaged thus resulting in a plan of poor utility. Therefore, instead, we optimally order the switching actions in the Stage-2 of the proposed approach.

The Stage-2 is also formulated as an MILP model to generate an optimal switching sequence over a horizon of ${T}$ time steps. The ${T}$ time steps are determined based on the total number of switching actions as obtained from the Stage-1 solution (${T}= \{t_{s1}, t_{s2}, t_{s3},...., t_{sn}\}$). To accompany for staged load picked up while including CLPU, each time step is further divided into 5 smaller steps. Although increasing the number of time steps will allow restoration of additional loads in the intermediate time steps, a higher value will increase the number of variables. Therefore, the time steps should be selected carefully based on how many switching actions are required from the Stage-1 solution.  

The problem variables and constraints include multiple copies of those from the Stage-1, one per time-step. The objective of this stage is to maximize the total load restored at each switching operation as defined in (16). This is to be accomplished without disrupting the service to already restored or uninterrupted customer loads. Constraint (\ref{stage2}) updates the positions of the line and load control switches at each time step while taking into account all possible closing/opening operations. 
\begin{itemize}
    \item Constraint (\ref{stage2}a) ensures that loads once picked are not interrupted during the consecutive sequence of restoration actions.
    \item Equation (\ref{stage2}b) ensures that the solution obtained in Stage-1 is guaranteed to be realized at the end of the switching sequence. That is, all loads that are energized as per the Stage-1 solution are picked up by the end of the restoration switching sequence.
    \item If any switch is operated once, it should not again change the status during the restoration. This is done to save the cost of switching operations. Thus, (\ref{stage2}c) and (\ref{stage2}d) restrict the reverse operation of switches during consecutive switching operations. 
    \item In Stage-2, we are required to operate one switch at a time. This is done to avoid the possibility of forming a loop configuration during switching operations. The constraint (\ref{stage2}e) enforces at most one switching operation per time step.
    \item Constraints (\ref{stage2}f) and (\ref{stage2}g) ensures that all sectionalizing-switches and tie-switches are in desired statuses as per the solution of Stage 1 in the final time-step of the switching operation. 
    \item The switches that do not take part in the restoration process are frozen at their original state for the entire duration of switching operation (See (\ref{stage2}h) - (\ref{stage2}j)).
\end{itemize}

\section{Results and Discussions}
The effectiveness of the proposed approach is validated using a multi-feeder 1069-bus test system consisting of four R3-12.47-2 PNNL taxonomy feeders connected using several tie switches \cite{kevin_PES2009}. The restoration problem is formulated as an MILP that can be solved using off-the-shelf solvers. The restoration formulation is developed in a python programming language where optimization for service restoration is modeled using PuLP modeling functions and solved using CPLEX 12.6 solver. The simulation is carried out on a PC with 3.4 GHz CPU and 16 GB RAM. Note that any effective solvers can be used to solve the problem.

 The taxonomy feeder R3-12.47-2 is a representation of a moderately populated urban area. The total load on the feeder is 4366.955 kW and 1299.206 kVAr. Four identical feeders are replicated to obtain the four-feeder 1069-bus distribution system where feeders are interconnected using seven normally open tie switches (see Fig. \ref{test_case}). 
 With a total of 1069 multi-phase physical buses (3444 single-phase buses), 152 sectionalizing switches, 190 possible cycles, and $122,586$ number of normal operational radial topologies the 1069-bus test case is a sufficiently large-scale model. We also incorporate several grid-forming utility-owned DGs in test case and the generation capacity limit of those DGs is shown in Fig. \ref{test_case}.
To ensure the ability to transfer the load to other feeders, the feeder loading is limited to 70\%, consequently, the feeder transformer capacity is 6.7 MVA. This system is assumed to be in a peak load condition. The allowable minimum single-phase voltage in this system is 259.698 V \cite{li2014distribution}. Thus, for a nominal voltage of 277.1 V, minimum voltage in p.u. is taken as 93.72\% (i.e., $V_{min}^\phi = 0.9372$ pu).

\vspace{-0.2 cm}
\subsection{Case Study}
Three different scenarios are simulated to validate the effectiveness of the proposed approach. For each study, the restoration is solved with and without DG. The Stage-1 solutions are presented in a tabular format and Stage-2 solutions are shown graphically to describe how the loads are picked up at each switching operation.

\begin{figure*}[t]
\centering
\begin{subfigure}{.45\textwidth}
  \centering
  \includegraphics[width=.95\linewidth]{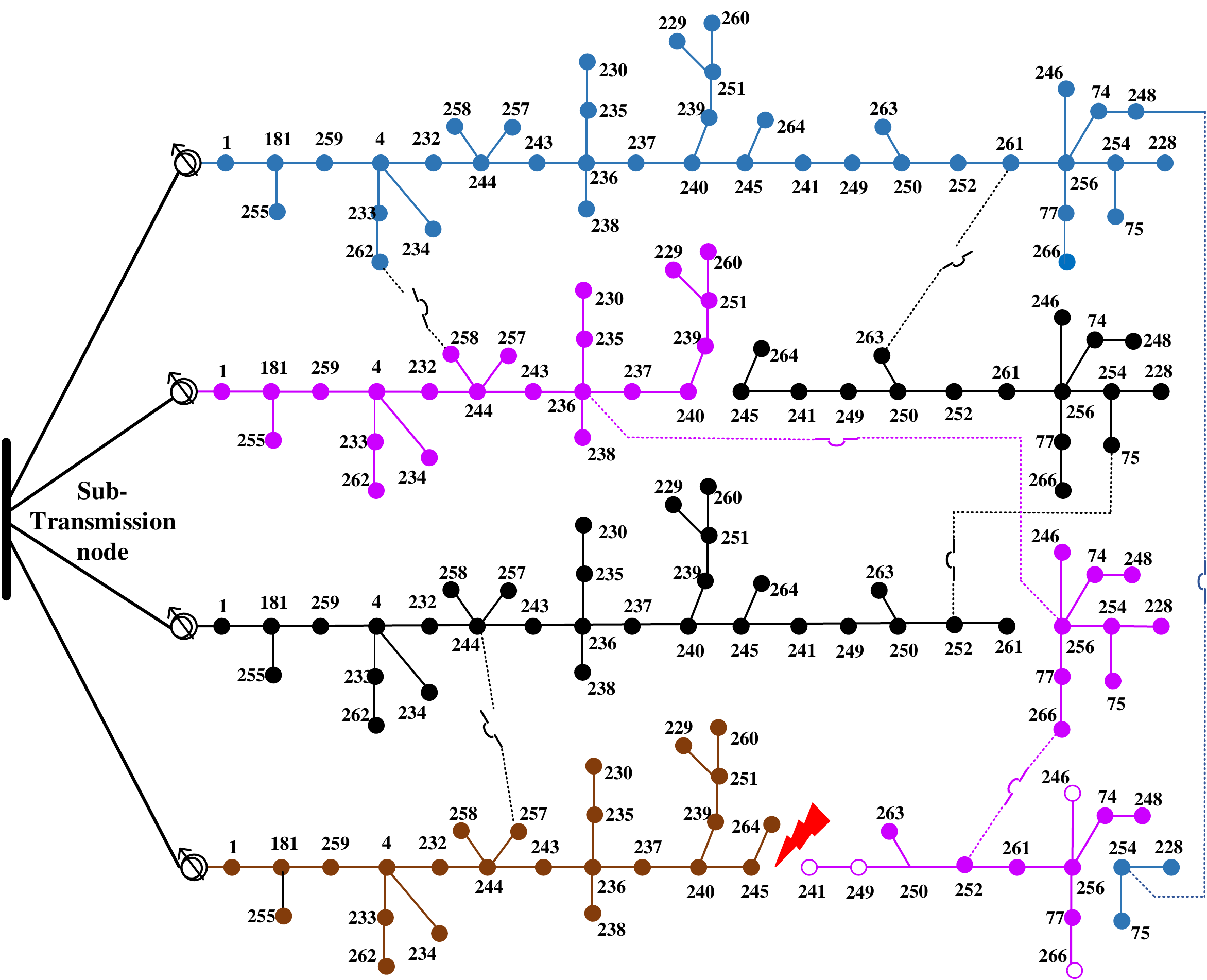}
  \caption{Circuit reconfiguration for Scenario 1}
  \label{fig:sub2_3}
\end{subfigure}
\begin{subfigure}{.45\textwidth}
  \centering
  \includegraphics[width=.95\linewidth]{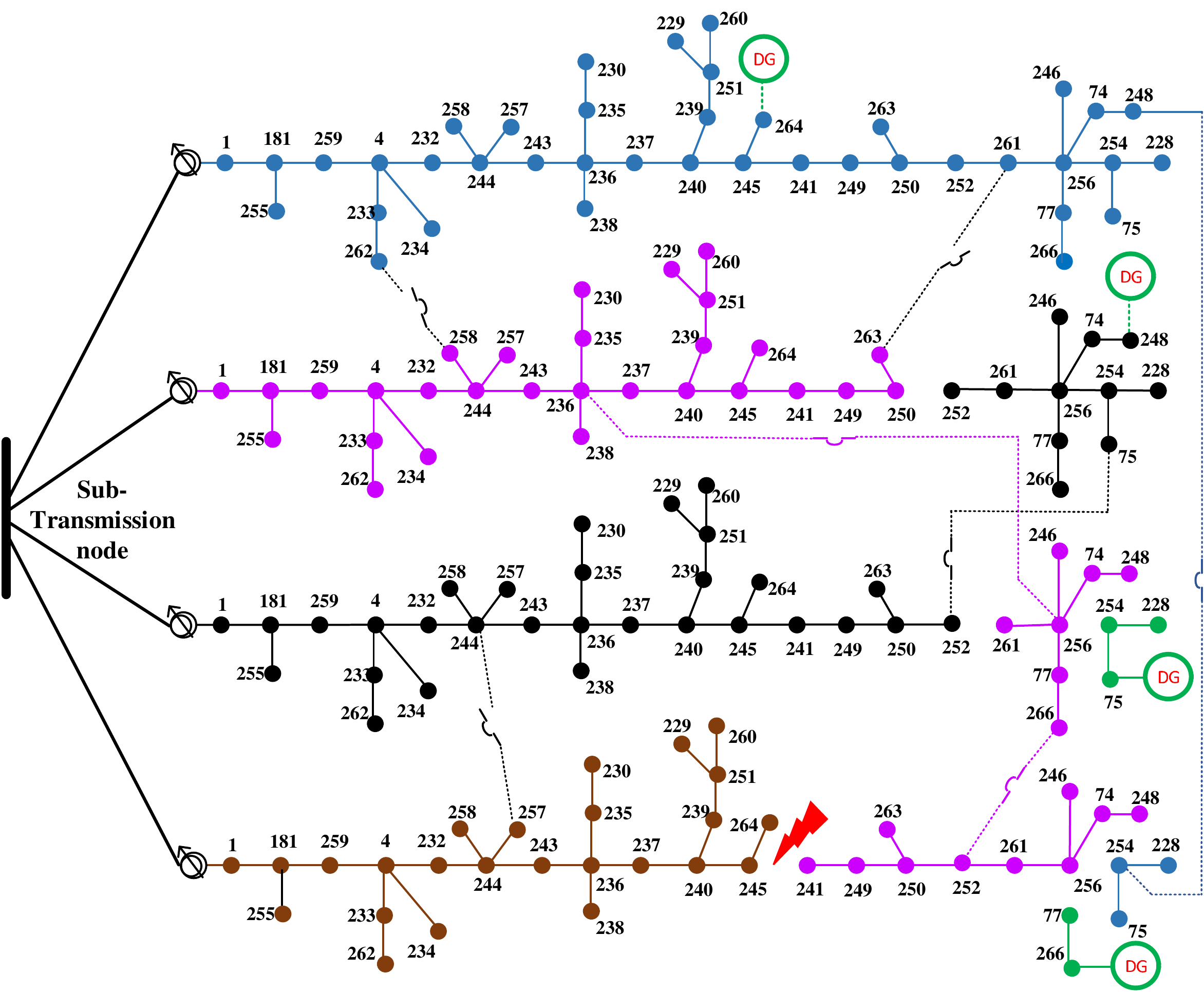}
  \caption{Circuit reconfiguration and DG islanding for Scenario 1}
  \label{fig:sub2_4}
\end{subfigure}
\caption{Restoration results with and without DGs. Different Color indicate the feeder reach for laods}
\label{fig:pictorially}
\end{figure*}

\begin{table*}[t]
    \centering
    \caption{ Restoration strategy for the four-feeder 1069-bus test case with DGs.}
    \label{case2}
    \vspace{0.2 cm}
    \begin{tabular}{c|c|c|c|c|c|c|c}
        \hline
            \multirow{2}{*}{Scenario} & \multirow{2}{*}{Line Fault} & Switch &Load Loss  &\multirow{2}{*}{ Isolation} & \multicolumn{2}{c|}{Switching Schemes without DGs}& Load Restored \\
            & & Tripped & kW& & Open & Close &kW\\
            \hline
             \multirow{6}{*}{1} & \multirow{6}{*}{241-159 (F-d)} & \multirow{6}{*}{7-241 (F-d)} &\multirow{6}{*}{3627.56}& \multirow{6}{*}{164-249 (F-d)}& \multirow{6}{5.7 em}{220-254 (F-d) 220-77 (F-d) 136-245 (F-b) 220-254 (F-c) 195-256 (F-c)} & \textbf{DG-75 $  $ (F-c)} &\multirow{6}{*}{3627.56}\\
            &&&&&&\textbf{DG-266 $  $ (F-d)}&\\
            &&&&&&75-252 (F-b and F-c) &\\
            &&&&&&248-254 (F-a and F-d)&\\
            &&&&&&266-252 (F-c and F-d)&\\
            &&&&&&236-256 (F-b and F-c) &\\
            \hline
             \multirow{5}{*}{2} & \multirow{5}{*}{181-182 (F-c)} & \multirow{5}{*}{267-1 (F-c)} &\multirow{5}{*}{4366.95}& \multirow{5}{*}{182-259 (F-c)}& \multirow{5}{5.7 em}{219-74 (F-a) 222-75 (F-b)  164-249 (F-b) 219-74 (F-b)} & \textbf{DG-248 $  $ (F-b)}   &\multirow{5}{*}{4366.95}\\
            &&&&&& 261-263 (F-a and F-b)&\\
            &&&&&& 248-254 (F-a and F-d)&\\
            &&&&&& 236-256 (F-b and F-c)&\\
             &&&&&& 75-252 (F-b and F-c)&\\
        \hline
        \multirow{4}{*}{3} & \multirow{4}{5.99 em}{193-195 (F-b) 105-191 (F-d) } & \multirow{4}{5.99 em}{223-261 (F-b) 223-261 (F-d) } & \multirow{4}{*}{6757.82}& \multirow{4}{5.7 em}{195-256 (F-b) 195-256 (F-d)}& \multirow{4}{5.7 em}{223-261 (F-c) 220-254 (F-d)} & \textbf{DG-266 (F-d)}&\multirow{4}{*}{6089.04}\\
        &&&&&&248-254 (F-a and F-d)&\\
        &&&&&&75-252 (F-b and F-c)&\\
        &&&&&&236-256 (F-b and F-c)&\\
        \hline
    \end{tabular}
\end{table*}

\subsubsection{Scenario 1}

In this scenario, a fault is simulated in the line 241-159 in F-d. As a result, an upstream switch 7-241 is tripped. Once the faulted zone is identified, an additional switch 164-249 in F-d is opened to isolate the fault from all possible directions; this is to ensure that during circuit reconfiguration the fault is not fed from any other feeder. After isolating the fault, it is observed that 3627.56 kW of load in F-d is disconnected. The first stage solution of the problem for this particular scenario is to open three sectionalizing switches and close four tie switches. The switch list generated by solving the Stage-1 of the problem is given in Table \ref{case1}. Next, the second stage of the problem is solved to generate the switching sequence based on the formulation detailed in Section 5. The sequence of operation is shown in Fig. \ref{fig:sc1}a. The switching actions show that a load transfer in feeder F-b and Feeder F-c is done to restore loads in F-d. Once the load transfer is done and switch between F-c and F-d is closed ($t_{s5}$), the CLPU effect is seen because loads in F-d have lost their diversity. However, after several time steps,  when loads under CLPU start gaining diversity, additional loads are picked up. The optimal solution for this case is to restore 2973.5 kW of outaged loads resulting in a load shedding of 654.04 kW. 

Next, the same scenario is simulated for the case that allows intentional islanding using grid-forming DGs. Unlike, the previous case, the Stage-1 solution restores all the outaged loads and requires a total of 11 switching operations including the formation of two DG-supplied islands (See Table \ref{case2}). The DGs located at F-c and F-d pick up loads and form separate islands to free the feeder-head transformer capacity to help restore additional loads. The sequence of operation is shown in Fig. \ref{fig:sc1}b where $5^{th}$ and $11^{th}$ switching instants correspond to the operation of DG switch. On incorporating DGs, the feeder is fully restored, however, at the cost of an increased number of switching operations. This is in agreement with the problem objective that prioritizes restoring customer loads over the number of switching operations. 

This overall scenario is pictorially represented in Fig. \ref{fig:pictorially} where the different colors of nodes indicate the feeder to which they belong. The solid and empty nodes signify whether loads in that zone are energized or not. During the normal operating condition, each feeder supplies its own load. For a given faulted scenario, feeders increase their boundary of operation and restore additional loads in the neighboring feeders using a suitable switching scheme (See Fig. \ref{fig:pictorially}a). Similarly, Fig. \ref{fig:pictorially}b shows the restoration solution with DG islands where all loads are picked up.

\subsubsection{Scenario 2}

\begin{figure*}[!t]
\centering
\begin{subfigure}{.49\textwidth}
  \centering
  \includegraphics[width=.99\linewidth]{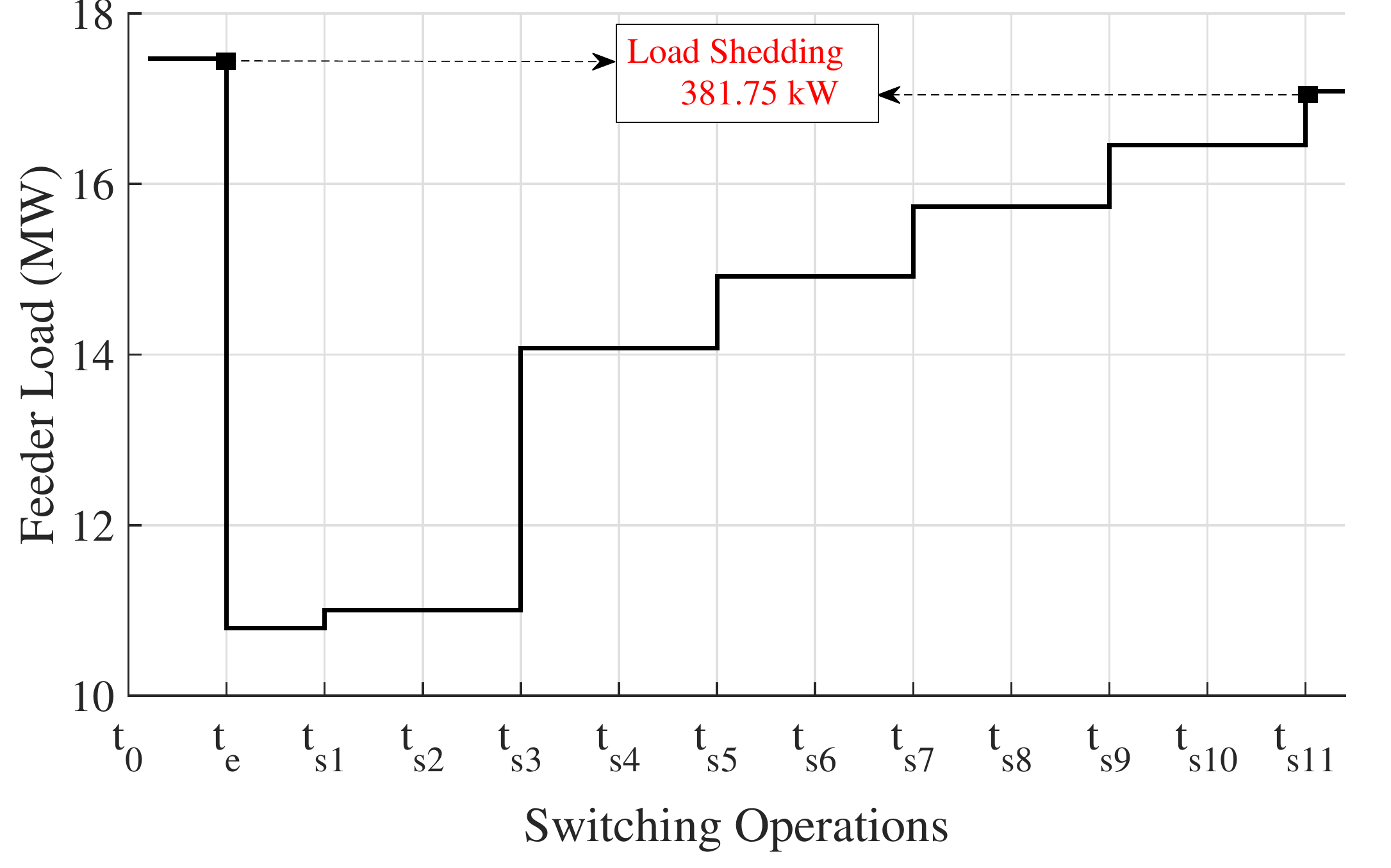}
  \caption{Feeder load picked up after sequence of switching operations.}
  \label{fig:sub2_5}
\end{subfigure}
\begin{subfigure}{.49\textwidth}
  \centering
  \includegraphics[width=.99\linewidth]{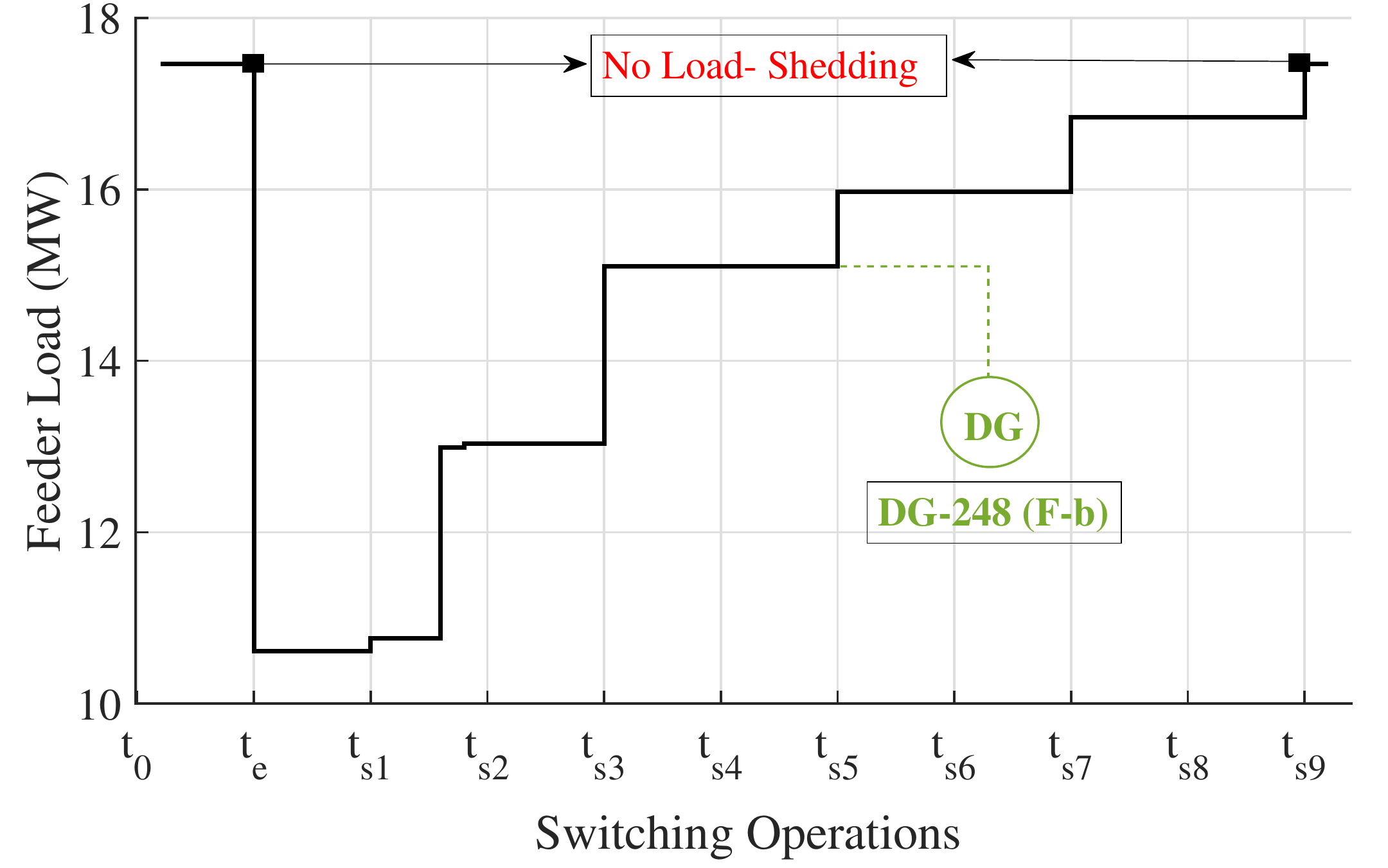}
  \caption{Feeder load picked up after sequence of switching operations including DG islanding. No load shedding}
  \label{fig:sub2_6}
\end{subfigure}
\caption{Switching sequence for Scenario 2 (Fault at Line 181-182 (F-c))}
\label{fig:sc2}
\end{figure*}

In this scenario, a fault in the line 181-182 in F-c is simulated. Because of fault, an upstream switch 267-1 is tripped. Once the faulted zone is identified, an additional switch 182-259 in F-c is opened to isolate the fault from all possible directions such that during the restoration, the fault is not fed from any of the healthy feeders. After isolating the fault, it is observed that 4366.95 kW of load in F-c is disconnected. The first stage solution to the problem for this particular scenario is to open 5 sectionalizing-switches and close 6 tie-switches. The switch list generated by solving the Stage-1 of the problem is given in Table \ref{case1}. Next, Stage-2 of the problem is solved to generate the switching sequence based on formulation detailed in Section 5. The sequence of operation is shown in Fig. \ref{fig:sc2}a. The switching actions show that the load transfer from feeder F-c to feeders F-a, F-b and F-d is done to restore the outaged loads in F-c. The optimal solution for this case is to restore 3985.20 kW of outaged load resulting in a load shedding of 381.75 kW. It is important to note that because of several switching actions, the load in feeder F-c load is picked up without observing any CLPU effect. Thus, one of the methods of address CLPU is to restore distribution feeders sequentially in several switching steps.

Next, the same scenario is simulated in the presence of DGs. Unlike, the previous case, the restoration solution restores all the outaged loads with a total of 9 switching operations and the formation of one DG-supplied intentional island (See Table \ref{case2}). The DGs located in feeder F-b picks-up the additional loads and forms a separate island. This frees up the feeder transformer to restore additional loads. The sequence of operation is shown in Fig. \ref{fig:sc2}b, where, $5^{th}$ switching instant corresponds to the operation of the DG switch. On incorporating DGs, the feeder is fully restored. However, the CLPU effect is observed here between the switching instances $s_1$ and $s_2$. 

\begin{figure*}[!t]
\centering
\begin{subfigure}{.47\textwidth}
  \centering
  \includegraphics[width=.99\linewidth]{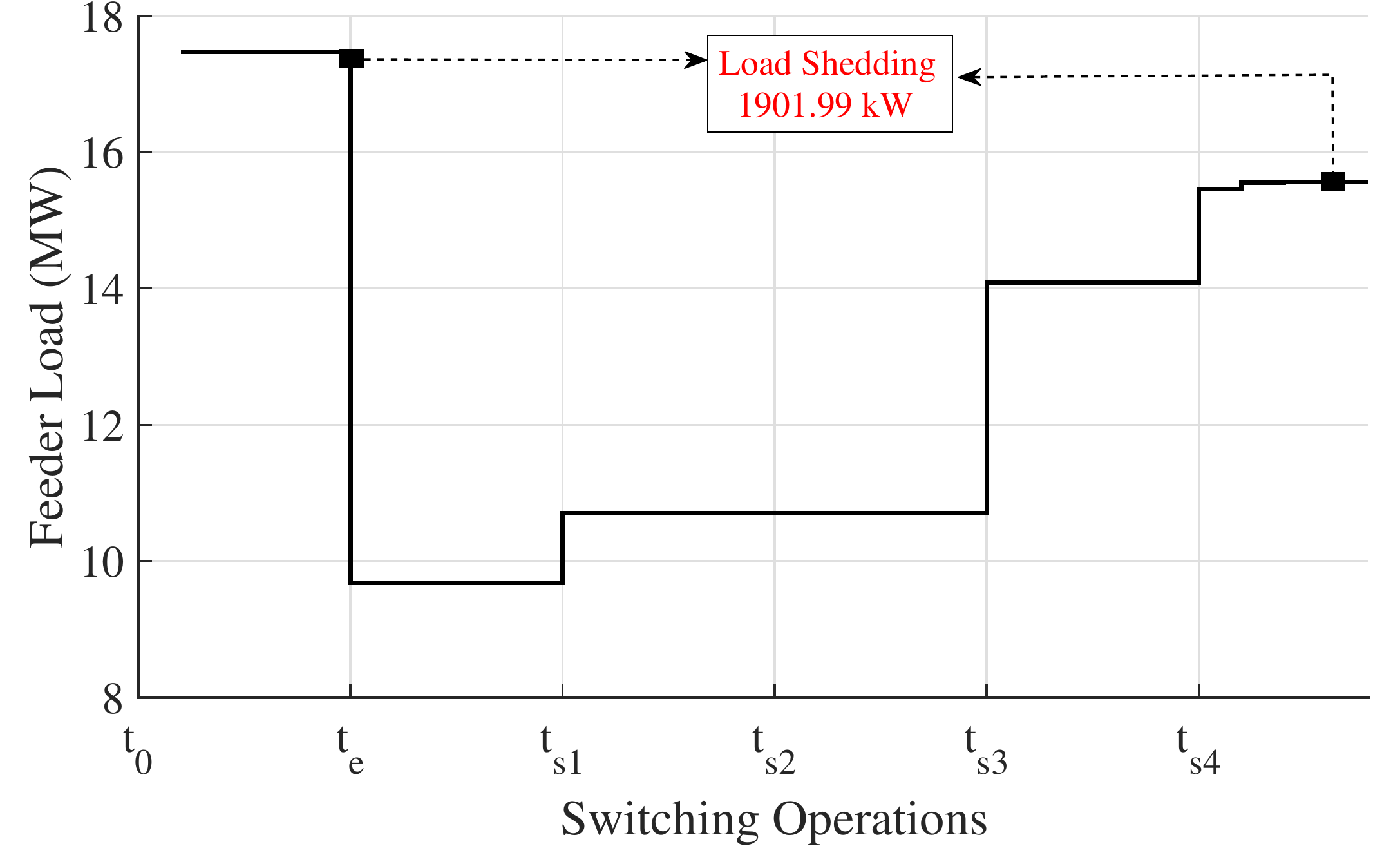}
  \caption{Feeder load picked up after sequence of switching operations.}
  \label{fig:sub2_7}
\end{subfigure}
\begin{subfigure}{.47\textwidth}
  \centering
  \includegraphics[width=.99\linewidth]{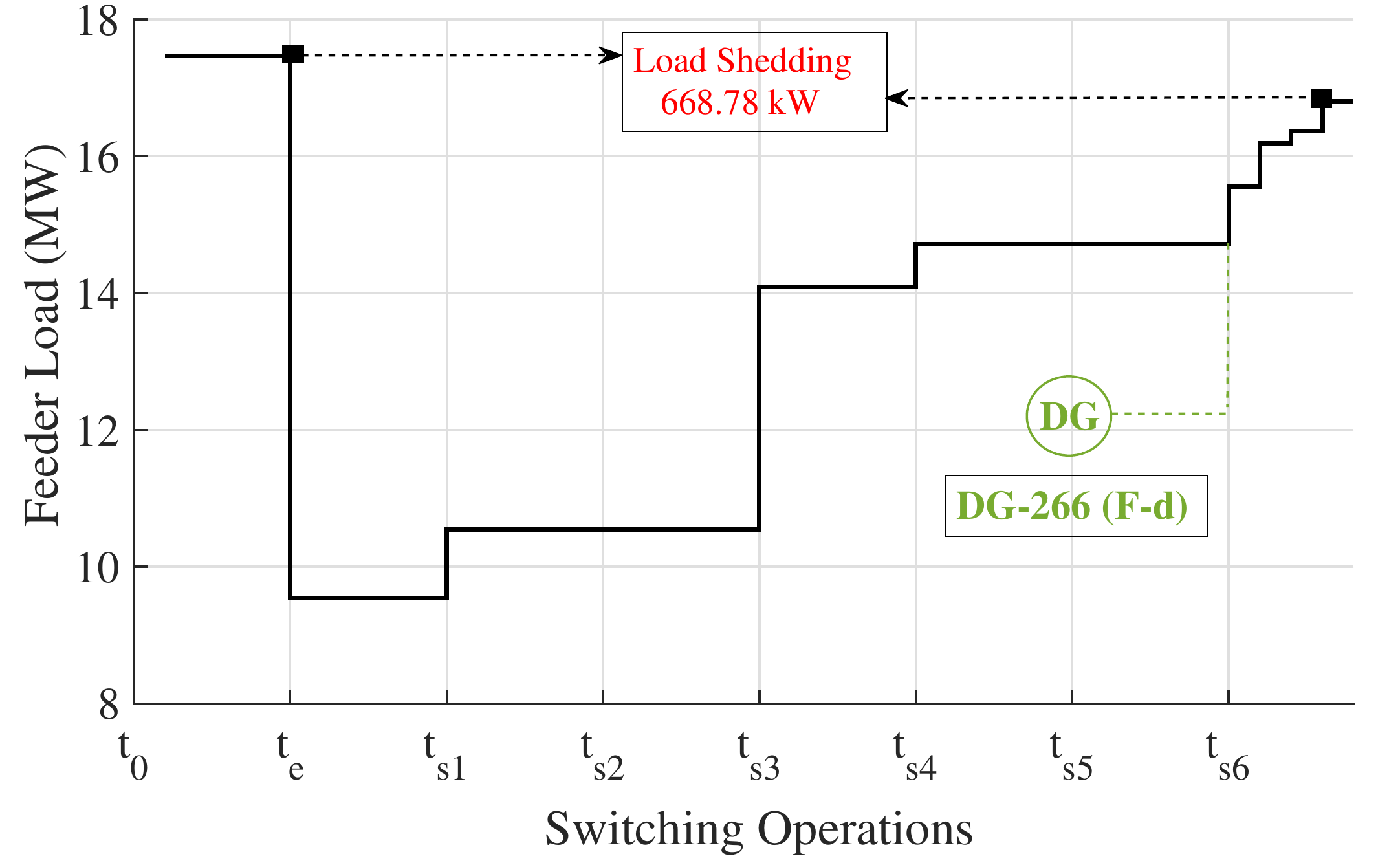}
  \caption{Feeder load picked up after sequence of switching operations including DG islanding. Load shedding reduced to 668.78 kW}
  \label{fig:sub2_8}
\end{subfigure}
\caption{Switching sequence for Scenario 3 (Fault at Line 193-195 (F-b) and Line 105-191 (F-d))}
\label{fig:sc3}

\end{figure*}

\subsubsection{Scenario 3}
In this scenario, a case with multiple faults is simulated. Similar to the previous two cases, the switches tripped because of faults. The switching operations required to isolate the corresponding faults are shown in Table \ref{case1}. After isolating the faults, it is observed that 6757.82 kW of the loads in feeders F-b and F-d are out-of-supply. The Stage-1 solution for this scenario is to open one sectionalizing switch and close three tie switches. The switch list is shown in Table \ref{case1}. Next, the Stage-2 of the problem is solved to generate the sequence of switching operation for the switch list that was obtained in Stage-1. The switching sequence is shown in Fig. \ref{fig:sc3}a where 6089.04 kW of the load is gradually picked up. 

Next, the same scenario is simulated in the presence of DGs. The number of switching operations is increased to six. The DG located in feeder F-d picks up additional loads and forms one island. This allows feeder F-d to pick-up additional loads thus reducing the total amount of load shedding from 1901.09 kW to 668.78 kW compared to the case without DGs. The sequence of operation is shown in Fig. \ref{fig:sc3}b. CLPU effect is seen when the DG is restoring the outaged loads during the last switching event. Note that the effect of CLPU is taken into account by operating the load control switch at different intermediate time-steps.

\begin{figure*}[t]
\centering
    \includegraphics[width=0.98\textwidth]{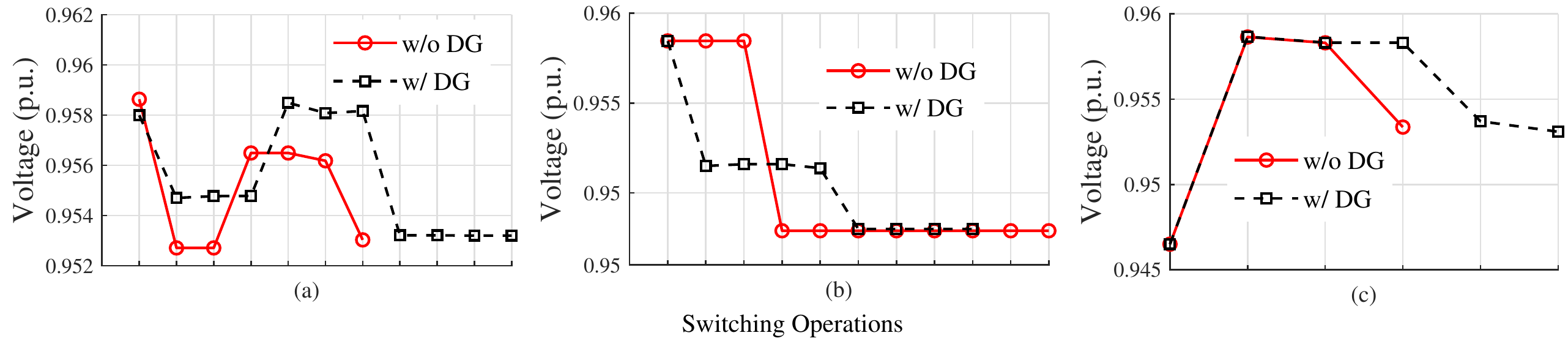}
    \caption{Voltage in p.u. at different switching instant for different scenarios. The grid (x-axis) in the plot corresponds to each switching instant.}
    \label{fig:v}
    \vspace{-0.2 cm}
\end{figure*}
\vspace{-0.3 cm}
\subsection{Computational Complexity}
 Table \ref{table:cc} shows the algorithm execution time for three different scenarios tested on the multi-feeder test case that is representative of a real-world distribution system. On an average, the proposed approach takes 26.01 seconds to solve the first stage and 256.95 seconds to solve the second stage of the problem.  
 \begin{table}[!h]
    \centering
    \caption{Simulation time for different cases}
    \label{table:cc}
    \begin{tabular}{c|c|c|c|c}
        \hline
        \hline
        \multirow{3}{*}{Scenario}&\multicolumn{4}{c}{Simulation time (Seconds)}\\
        &\multicolumn{2}{c|}{With DG}&\multicolumn{2}{c}{Without DG}\\
        &Stage 1 & Stage 2&Stage 1 & Stage 2\\
        \hline
        1 & 37.25 & 410.23 & 10.83 & 320.45  \\
        2 & 38.73 & 365.56& 35.25  & 240.56 \\
        3 & 9.70 & 261.41 & 24.31 & 43.45 \\
        \hline
    \end{tabular}
\end{table}

The ability to scale a research project to a commercial deployment successfully is a critical piece that the methods present in the existing literature overlook. The real-world power distribution network has thousands of buses and lines with several possible combinations for reconfiguration. Thus, the number of possible topologies for the multi-stage process is significantly high. Along these lines, a non-linear problem formulation is known to not scale for the multi-stage formulation to generate a sequence of switching actions for restoration for a real-world distribution system. Similarly, although heuristic-based methods will always result in an optimal restoration plan, they require an exhaustive search of all possible options at each switching instant thus limiting their applicability for large distribution systems with numerous options for reconfiguration. Given the existing methods, the proposed MILP model is scalable for the real-world distribution system and provides an operationally feasible restoration plan along with the switching sequence in a computationally tractable manner.

\subsection{Validation of Linear Power Flow}
Since the formulation discussed in Section 4 approximates power flow equations, the results obtained from the proposed linear and nonlinear power flow models are compared with the power flow solution obtained using OpenDSS, a distribution system simulator that solves the nonlinear power flow model. The largest errors in apparent power flow and bus voltages are reported for the R3-12.47-2 in Table \ref{lnl} for different loading conditions. Note that the three-phase linear power model is sufficiently accurate in modeling power flow equations for an unbalanced system. Since the losses are ignored in flow equations (equations \ref{pflow}a and \ref{pflow}b), the linear model incurs higher error in flow quantities. However, since the voltage drop due to flow quantities is included in the linear model (equation (\ref{pflow}c) and (\ref{pflow}d)), the bus voltages are well approximated.
\begin{table}[h]
    \centering
    \caption{Comparison of linear power flow formulations against OpenDSS}
    \label{lnl}
    \begin{tabular}{ccccc}
    \hline
        Feeder&\% Loading & $S_{flow} (\%)$ & $V(p.u.)$\\
        \hline
        \multirow{2}{*}{R3-12.47-2}&75 &3.24 &0.001 \\
        &100 &2.84  &0.002 \\
         \hline
    \end{tabular}
\end{table}

In addition to the verification of approximate power flow formulations, an exact power flow analysis is done for each scenario in OpenDSS. Each tree is simulated in OpenDSS with a detailed line and load model. As mentioned in Section 5.2, each switching instant should be operationally feasible, OpenDSS model is modified accordingly to simulated the detailed model of the distribution network at each switching instant. The actual power flow results are obtained and the minimum node voltage is reported (See Fig. \ref{fig:v}). It is observed that for each scenario, the minimal node voltage is within the specified limit at every switching instant.

\subsection{Comparison with the State-of-the-art}
In this sub-section, the proposed approach is compared with two state-of-art methods for DSR from references \cite{li2014distribution} and \cite{thiebaux2013planning}. The approach presented in \cite{li2014distribution} is based on spanning tree search method while \cite{thiebaux2013planning} presents a MILP model for the service restoration.

The approach detailed in \cite{li2014distribution} applies a spanning tree search algorithm to generate all possible network topologies to restore loads in an outaged area. Several candidate network topologies are generated based on possible switching operations and for each topology, a power flow feasibility analysis is done. An optimal network topology that satisfies the operating constraints is identified next and appropriate switches are opened/closed to reach the desired network configuration. Note that the approach is based on selective cyclic interchange operation which is borrowed from \cite{mayeda1965generation}. Given an operating tree $\mathcal{T}_o$, some other trees can be obtained by replacing an edge of tree with another edge in the fundamental cutset $S_e(\mathcal{T}_0)$. The set of distinct trees is given by (\ref{cutset}). 
\begin{equation}\label{cutset}
    \big\{t|t = \mathcal{T}_o \oplus \{e,e_i\}, e_i \in S_e(\mathcal{T}_0), e_i\neq e\big\}.
\end{equation}
where, $S_e(\mathcal{T}_0)$ is the fundamental cutset defined by edge $e \in \mathcal{T}_o$. 

The spanning tree-based approach relies on an exhaustive search of all possible spanning trees. It then stores the list of all spanning trees and for a given outage scenario selects the optimal restoration topology from the stored list. First, as the system complexity grows with the integration of additional grid-forming DGs and switches, searching and storing all possible spanning trees will become computationally expensive. Further, the stored set of spanning trees need updating every time a new resource that can help with restoration is added to the system. Further, the spanning tree-based approach in its current form as proposed in \cite{li2014distribution} only works for a single fault scenario. This is because a separate repository of spanning trees needs to be stored as multiple fault scenarios that can be cumbersome as the combinations for multiple faults grow. This further complicates the implementation of the spanning tree-based algorithms for a practical distribution network. On the contrary, the approach proposed in this paper dynamically obtains the optimal network configuration using mathematical optimization techniques for any number of faults seen by the underlying system. Further, the proposed MILP model can be easily expanded to included new devices (DGs and switches).

\begin{figure}[t]
    \centering
    \includegraphics[width=0.45\textwidth]{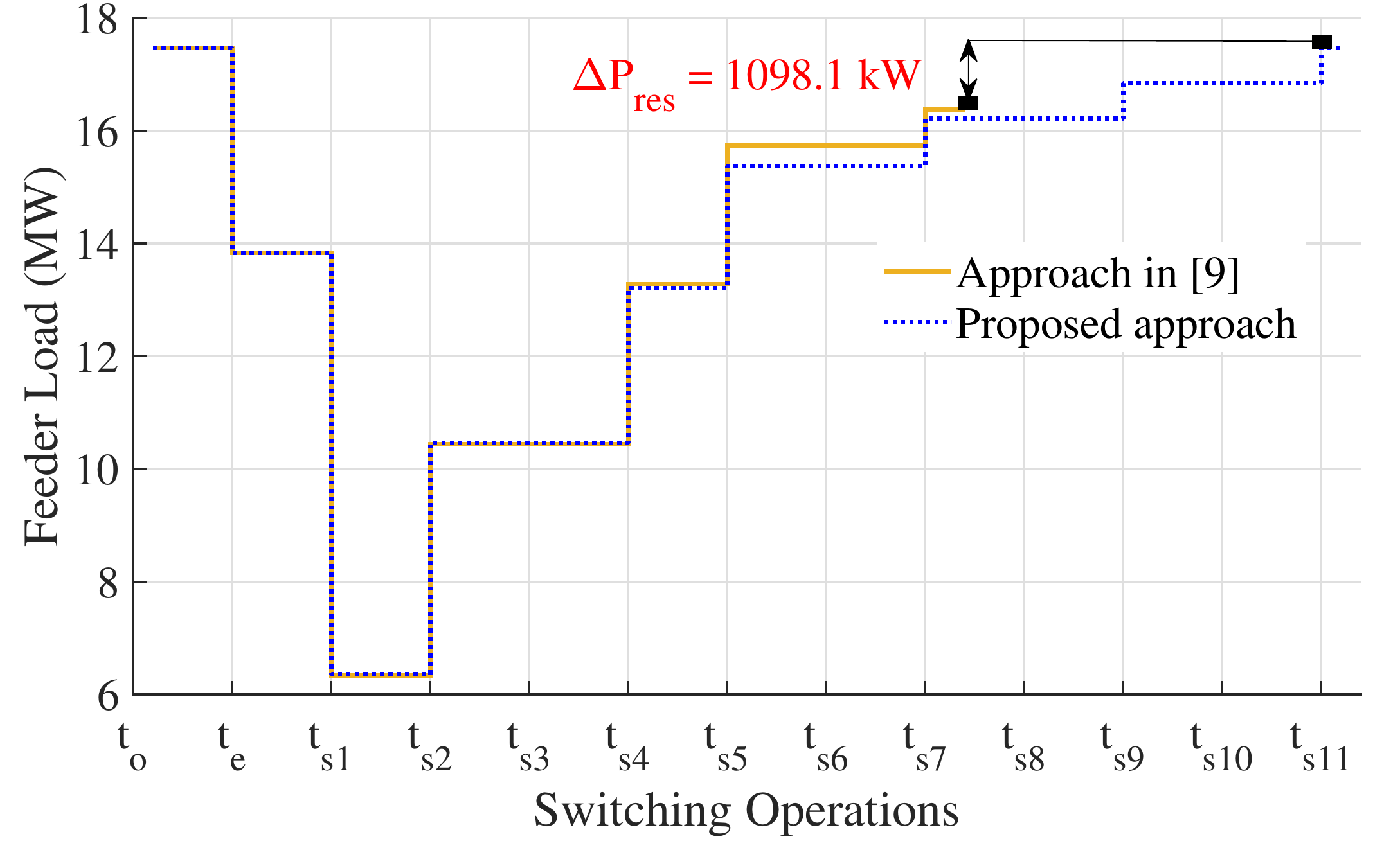}
    \caption{Comparison of total load picked up for Scenario 1.}
    \label{fig:my_label}
\end{figure}

\begin{table}[t]
    \centering
    \caption{Comparison of proposed approach with  \cite{li2014distribution} and \cite{thiebaux2013planning}  }
    \label{comp}
    \begin{tabular}{l|c|c|c}
    \hline
        \multirow{2}{*}{Attributes} & Proposed &  \multirow{1}{*}{Mathematical }  &  \multirow{1}{*}{Graph-theoretic} \\
        &approach&optimization \cite{thiebaux2013planning}&\cite{li2014distribution}\\
        \hline
        Methodology & MILP & MILP & Spanning tree \\
        Test System Size & 1069-bus & 207-bus & 1069-bus\\
        Multiple faults & \checkmark & \checkmark & \Large{$\times$}\\
        DG-islanding & \checkmark & \Large{$\times$} & \checkmark\\
        CLPU effects & \checkmark & \Large{$\times$} & \Large{$\times$}\\
        \hline
    \end{tabular}
    \vspace{-0.3 cm}
\end{table}

The approach detailed in \cite{thiebaux2013planning} presents an MILP framework for DSR for switching sequence generation. However, the method proposed in \cite{thiebaux2013planning}  does not take into account the DG-assisted restoration and ignores the loss in load diversity after an outage due to the CLPU effect. This may lead to sub-optimal restoration schemes especially when the CLPU effect is present and when DG can be used to pick-up additional loads. We simulated Scenario \#1 using the method proposed in \cite{thiebaux2013planning}. The difference in the solutions for load restoration using our approach and the one proposed in \cite{thiebaux2013planning} is shown in Fig. \ref{comp}. It is observed that our approach is better able to restore the distribution system and restores larger MWs of load at the cost of increased switching operations. This is per the objective function where the primary objective is to restore as much load as possible. Since the formulation in \cite{thiebaux2013planning} ignores the loss in load diversity after an outage, the size of load shedding is 1098.1 kW ($\Delta P_{res}$) as opposed to 654.04 kW as per our approach (See Fig. \ref{fig:sc1}a). This is because the loads are underestimated while formulating the problem in  \cite{thiebaux2013planning} and operational constraints are violated thus requiring additional loads to be shed. 

The overall comparison of the proposed approach against state-of-art methods in \cite{li2014distribution} and \cite{thiebaux2013planning} is summarized in Table \ref{comp}.

\vspace{-0.5 cm}

\section{Summary}
In this paper, we proposed a generalized framework for service restoration for an unbalanced three-phase power distribution system that uses all available resources (healthy feeders and DGs) for maximizing the total restored loads.  A two-stage framework is proposed where (1) Stage-1 generates the switching operations required to restore a maximum amount of customer loads using all available resources, and (2) Stage-2 generates the sequence of switching actions that help faulted system in transitioning from the post-fault condition to restored configuration without violation feeder's operating constraints during consecutive switching operations. The stage-2 takes the loss in load diversity due to cold-load pick-up (CPLU) into account and coordinates the feeder reconfiguration and active DG islanding under CLPU conditions. The simulation results demonstrate that grid-forming DGs help restores additional loads for the distribution system. Using the Stage-2 solution, one can energize the distribution system sequentially without violating the operational and topological constraints. The results indicate that our approach can find optimal switching sequences for a large unbalanced distribution system with thousands of buses on an average within 5 mins. This helps the operator to safely execute the restoration solution promptly.

Finally, it is worth mentioning that reliability indices such as System Average Interruption Duration Index (SAIDI), Customer Minutes of Interruptions (CMI), and Customer Average Interruption Duration Index (CAIDI) depend on restoration time after the fault has occurred. Therefore, with the execution of the restoration solution generated by the proposed approach, the utility companies can quickly restore the faulted feeders that can help significantly improve the reliability indices. 

\bibliographystyle{ieeetr}
\balance
\bibliography{references}

\end{document}